\newcommand{\cal}{\mathcal}
\newcommand{\ud}{\mathrm{d}}

\documentclass[twocolumn,psfig,showpacs,preprintnumbers,amsmath,amssymb]{revtex4}
\usepackage{graphicx}	
\usepackage{dcolumn}	
\usepackage{bm}		
\usepackage{amsmath}
\usepackage{amssymb}
\usepackage{amsfonts}
\usepackage[T1]{fontenc}
\usepackage{exscale}
\usepackage{epsfig}
\usepackage{portland}
\usepackage{changebar}
\usepackage{epic}
\usepackage{color}
\usepackage{rotating}
\usepackage{eepic}
\begin{document}
\preprint{GSI-2004-13, IPNO-DR-04-03}
\title{
	High-resolution velocity measurements on 
	fully identified light nuclides\\ produced 
	in $^{56}$Fe + hydrogen and $^{56}$Fe + titanium systems
%
}
%
%
\author{P. Napolitani$^{1,2,}\footnote{This work forms part of the
		PhD thesis of P. Napolitani, Universit\'e Paris XI, Paris 2004, 
		n.7620, preprint IPNO-T-04-14}$, 
	K.-H. Schmidt$^{2}$, 
	A.S. Botvina$^{2,3}$,
	F. Rejmund$^{1}\footnote{On departure to 
GANIL, Blvd. H. Becquerel, B.P. 5027, 14076 Caen, France}$, 
	L. Tassan-Got$^{1}$ and 
	C. Villagrasa$^{4}$}
\affiliation{$^{1}$~IPN Orsay, IN2P3, 91406 Orsay, France}	
\affiliation{$^{2}$~GSI, Planckstr. 1, 64291 Darmstadt, Germany}
\affiliation{$^{3}$~Inst. for Nuclear Research, Russian Academy of Sciences,
	117312 Moscow, Russia}
\affiliation{$^{4}$~DAPNIA/SPhN CEA/Saclay, 91191 Gif sur Yvette, France}
%
%
\date{\today}
%
%
\begin{abstract}
	New experimental results on the kinematics and the residue
production are obtained for the interactions of $^{56}$Fe projectiles
with protons and $^{\mathrm{nat}}$Ti target nuclei, respectively, at the
incident energy of 1 $A$ GeV.
	The titanium-induced reaction serves as a reference case for
multifragmentation.	
	Already in the proton-induced reaction, the characteristics
of the isotopic cross sections and the shapes of the velocity spectra
of light residues indicate that high thermal energy is
deposited in the system during the collision.
	In the $^{56}$Fe$+p$ system the high excitation seems to favour
the onset of fast break-up decays dominated by very asymmetric
partitions of the disassembling system.
	This configuration leads to the simultaneous formation of one
or more light fragments together with one heavy residue.
\end{abstract}
%
%
%
%
 \maketitle
%
%
%
%
\section{
	Introduction
}
%
	For the last decades, the investigation of the maximum
excitation energy that a nuclear system can hold has remained as much a
challenge as the description of the decay of a hot collision remnant,
excited beyond the limits of nuclear binding.
	It is commonly assumed that other decay modes than
fission and evaporation	prevail at high excitation energy.
	These modes are often described as a simultaneous
break-up of the hot system in many parts, named ``multifragmentation''.
	The excitation energy above which multifragmentation appears
is still a source of intense theoretical and experimental research.
	A point of particular interest is to recognise the distinguishing
traits denoting this decay mode when the excitation is just
sufficient for its onset.
	In line with this investigation, one foremost aspect of
intense discussion is the connection of the kinematics of the residues
to the kind of equilibration process involved in the
earliest stages of
the decay.
	This question is related to the complementary effort in
constructing physical models to deduce the formation cross sections
of the residues when the excitation energy of the system is taken as
initial condition.
	Especially light residues are suited for this purpose.
	Several details of the deexcitation mechanism could emerge
from the kinematics of light fragments, due to the high sensitivity in
probing the Coulomb field of the decaying system.
	Moreover, the distribution of their isotopic
cross sections carry additional signatures connected to
different decay modes.

	The experimental data exploited in this work are part of a vast 
experimental campaign devoted to the collection of nuclear data for the 
design of accelerator-driven subcritical reactors~\cite{Villagrasa03} and to the 
investigation of spallation reactions in the cosmos~\cite{George01}. 
	The full set of production cross sections and fragment mean recoil
velocities measured in this framework for the reaction 
$^{56}$Fe$+p$ and $^{56}$Fe$+^{\mathrm{nat}}$Ti at energies ranging from
$300$ to $1500$ MeV is presented elsewhere~\cite{Villagrasa04}.
%
\subsection{
	The formation of light residues
}
%
	Light residues can be generated in several kinds of 
processes. 
	One of these, the binary decay of an excited greatly 
thermalised complex, named compound nucleus, was widely 
studied~\cite{Sanders99}. 
	We might also recall that evaporation of nucleons
and light nuclei and symmetric fission are just the opposite
extremes of the manifestation of this process: there is a gradual
transition from very asymmetric to symmetric configurations
in the division of decaying compound nuclei,
and thus all binary decays of a greatly thermalised
system can be named fission in a generalized sense.
This generalization was introduced by
Moretto~\cite{Moretto75,Moretto89}.
	A compound system far below the
Businaro-Gallone point~\cite{Businaro55,Businaro55bis}
(like iron-like nuclei) undergoes very
asymmetric fission, resulting in a characteristic
U-shape in the mass distribution of the yields.
	A minimum located at symmetry in the yield
mass spectrum corresponds to a maximum placed at
symmetry in the ridge lines of the potential.
	In configurations where a heavy partner is present,
the whole decay process is dominated by the binary decay, and an
additional evaporation of single nucleons would not disturb the
kinematics remarkably.
	Such a process exhibits the
typical feature of the population of the shell of a
sphere in velocity space, in the reference frame of the mother
nucleus.

	At high excitation, multifragmentation becomes the
competing process to compound-nucleus reactions.
	There is a fundamental difference between the
binary decay of a compound nucleus and the simultaneous
disintegration of a hot collision remnant in several
constituents.
	The difference is in the kind of instabilities
which are the reason for the decay, and is reflected in
the kind and in the time evolution of the consequent
equilibration process followed by the system.	
	
	A hot nucleus with an excitation energy above
the threshold for emission of particles or clusters
(including fission) has the possibility to decay by any
of the open channels.
	If the excited system is not too hot,
the favoured process is a reordering of its
configurations: a great number of arrangements are
available where all nucleons remain in states below the
continuum, occupying excited single-particle levels
around the Fermi surface.
	Oscillations in fission direction are included
in this picture as well, but too rarely the fission
barrier is reached.
	Rather seldom, compared with this thermal
chaotic motion of the system, one nucleon acquires
enough energy to pass above the continuum and may
eventually leave the nucleus.
	This picture might be extended to cluster decay and to
fission.
	Since this decay is a rare process, one
evaporation event, or fission event, proceeds after the
other, sequentially.
	In this process, the compound system follows a
dynamic trajectory in deformation space, which is
governed by the potential-energy surface and the
dynamic properties of the compound system, related,
for instance, to the inertia tensor and dissipation tensor.
	All decays are binary.
	 	
%
%
\begin{figure}[b!]
\begin{center}
\includegraphics[width=\columnwidth]{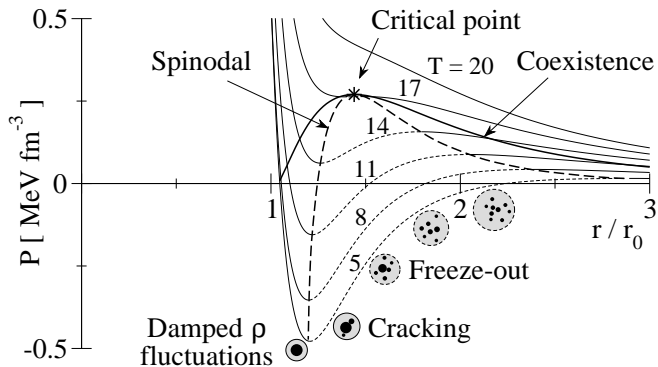}
\end{center}
\caption
{
	Idealistic plot of the phase diagram of nuclear matter,
deduced from a Skyrme force (\cite{Jaqaman83} parameterised according 
to~\cite{Levit85}).
	Pressure is shown as a function of the average relative 
nucleon distance r normalized to the distance $r_0$ at ground state.
	System configurations are drawn as possible final results 
of the expansion phase. When the thermalization path leads to the
coexistence region, out of the spinodal region, damped density 
fluctuations occur. 
	In the spinodal region density fluctuations are unstable
and lead to cracking.
	At low density freeze-out is attained with different possible
partition configurations: fragments are free to leave the system.
}
\label{fig:fig1}
\end{figure}
	If the system becomes drastically more
unstable, this picture is not valid anymore.
	The exploration of possible states of the
excited system includes numerous unstable
configurations.
	Thus, the disintegration can not be understood
as a sequence of binary decays, but rather portrayed as
a simultaneous break-up in several
constituents~\cite{Randrup81,Gross97,Botvina85,Bondorf85,Bondorf95}.
	The disintegration is simultaneous in the sense
that it evolves in so short a time interval
($10^{-22}$-$10^{-21}$s) that the ejected fragments can still
exchange mutual interactions during their
acceleration in the Coulomb field of the system.	
	In heavy-ion collisions, part of the
excitation could be introduced in the system in the form of
compressional energy. According to the impact parameter and
the incident energy, the interaction might result in a
very complex interplay between dynamic effects
(beside compression, also deformation and rotation degrees of
freedom) and thermal excitation.
	This is the case of central collisions in the
Fermi-energy range.
	On the contrary, peripheral heavy-ion collisions at
relativistic energies may be rather pictured according to an
``abrasion'' process~\cite{Gaimard91,Brohm94}, where the remnant is 
formed by the spectator nucleons, heated by mainly thermal energy.
	In this case, the role of compressional energy has
minor incidence.
	Even with proton projectiles, the multifragmentation regime
might be accessible when very high excitation is introduced in the
nucleus.
	In reactions induced by relativistic protons (but also by
very light nuclei), the dynamic effects of the collision have even
smaller importance.
	The excitation energy is almost purely thermal.
	Some authors even attributed the specific name of
``thermal multifragmentation'' to this particular process
(see the review articles~\cite{Karnaukhov99,Karnaukhov03}).
	It might be suggested that proton-induced relativistic
collisions are better suited than ion-ion collisions for
investigating thermal properties of nuclear matter
(e.g. \cite{Karnaukhov99,Hirsch84,Andronenko86,Kotov95,Avdeyev98}).
	In finite nuclei, the transition from the fission-evaporation mode to
multifragmentation manifests rather smoothly.
	This opening of break-up channels even inspired
interpretations in line with the liquid-gas phase transition of
nuclear matter~\cite{Richert01,Chomaz04,Pochodzalla95,Borderie02}.
	The similarity of the nucleon-nucleon interaction with
the Lennard-Jones molecular potential suggests that infinite neutral
nuclear matter resembles a Van-der-Waals fluid~\cite{Sauer76}.
	As shown in fig.~\ref{fig:fig1}, also in the phase diagram 
of nuclear matter an area of liquid-gas coexistence can be defined.
	In this region, the ``dense'' phase of nuclear droplets is in
equilibrium with the ``gaseous'' phase of free nucleons and light
complex particles.
	Within the Hartree-Fock approximation, according to the type
of Skyrme force chosen for obtaining the nuclear equation of state,
the critical temperature $T_c$ was calculated to vary in a range of around
15 to 20 MeV for nuclear matter~\cite{Sauer76,Jaqaman83,Levit85}.
(One of the latest investigations, based on an improved Fisher's 
model~\cite{Elliott02} indicated $T_c=6.7\pm0.2MeV$ for finite nuclear systems.
	This value is source of controversy, 
e.g.~\cite{Karnaukhov03,Karnaukhov03bis}).
	During the reaction process, the system explores different
regions of the phase diagram.
	Since at relativistic energies the collision is related to short
wavelengths, the hot remnant should reach high positive
values of pressure $P$ due to thermal energy (rather than mechanical
compression, characteristic of Fermi-energy collisions) without
deviating sensibly from the initial density $\rho_0$.
	It is commonly assumed that at this stage the system is
still not thermalised and it undergoes expansion in order to attain
equilibrium
(There exist also opposite interpretations assuming
thermalization already before expansion and a successive
``Big-Bang-like'' expansion out of equilibrium~\cite{Campi03}).
	If the initial pressure is high enough, the subsequent
expansion could lead to rather low densities, and the system, 
after dissipating the incoming momentum, could reach a point belonging 
to the spinodal region.
	Due to the inverse relation between pressure and density
$dP/d\rho<0$, this region is unstable, and density fluctuations are
magnified.
	The nucleus breaks apart due to spinodal instability.
	The system disassembles also due to Coulomb instability.
	The inclusion of the long-range Coulomb interaction in the
equation of state was introduced by Levit and
Bonche~\cite{Levit85}, with the result that the solution of the
coexistence equation vanishes above a ``limiting temperature''
$T_{lim}$, in general much lower than $T_c$, depending on the
conditions taken for the calculation
(see also \cite{Jaqaman89,Jaqaman89bis}).
	Density fluctuations reflect a continuous evolution of
the size and number of nuclear droplets from a configuration to
another~\cite{Bugaev01}.
	If the average mutual distance among the nucleons exceeds
the strong nuclear interaction range (i.e. about
$\sqrt{<\sigma_n>/\pi}$, where $<\sigma_n>$ is the average
nucleon-nucleon collision cross section), the break-up configuration
``freezes'' and the formed nuclei and nucleons fly away freely,
all carrying signatures of the so-called freeze-out temperature of
their common source.
	From comparing results from different experimental
approaches e.g.~\cite{Hirsch84,Pochodzalla95,Schmidt02,Napolitani02}
this temperature is found to be restricted to a range of 5 to 6 MeV
(corresponding to a range of excitation energy per nucleon around 2.5 to
3.5 MeV), quite independently of the reaction.
	This finding, not directly compatible with the phase diagram
of ideal nuclear matter even suggested to search for a
``characteristic temperature'' of fragmentation~\cite{Friedman88}.
	The break-up configuration at freeze-out is expected to
reflect the excitation energy of the system.
	The dense phase of highly heated systems should have the
aspect of an ensemble of copious almost-equal-size light fragments.
	At reduced excitation, just sufficient for attaining the
freeze-out, the break-up partition might evolve to more asymmetric
configurations, where the formation of a heavy fragment close to
the mass of the hot remnant is accompanied by one or more light
fragments and clusters.
	As an extreme, this configuration might even reduce to a
binary asymmetric decay.
	In the case of a very asymmetric split of the system,
the partition multiplicity has minor influence on the kinematics
of the light ejectiles.
	The emission of light particles populates spherical shells
in velocity space and can not be easily distinguished by the
kinematics from a binary decay when large mass-asymmetries
characterize the partition.
	A binary or binary-like decay issued from a break-up
configuration is a ``fast'' process.
	Compared to asymmetric fission, asymmetric break-up
decays should result in a similar U-shape of the mass
spectra of the yields.
	On the other hand, break-up decays should be reflected in
the higher magnitude of the yields, and in the emission kinematics
that, still mostly governed by the Coulomb field, should exhibit an
additional contribution due to the eventual expansion of the source.
%
\subsection{
	Measurement of light-fragment properties
}
%
	Great part of the information on
light-particle emission at high excitation energies
was collected in 4-$\pi$-type experiments, suited for measuring the
multiplicity and the correlations of intermediate-mass
fragments~\cite{ALADIN_MSU,INDRA,EOS}.
	Still, the measurement of correlations and the
linear-momentum-transfer was the basis for pursuing intense
researches on the transition from the formation of compound
nuclei to multifragmentation~\cite{Klotz-Engmann87,Klotz-Engmann89}.
	
	In this work, we discuss additional results derived
from new inclusive measurements of the reactions $^{56}$Fe$+p$
and $^{56}$Fe+$^{\mathrm{nat}}$Ti at 1 $A$ GeV, effectuated in inverse
kinematics with the FRagment Separator (FRS)~\cite{FRS} at
GSI (Darmstadt).
	The experimental set-up was not intended to measure
multiplicity and correlations, but to provide formation
cross sections and high-resolution velocity spectra for
isotopically identified projectile-like residues.
	The excitation of the $^{56}$Fe$+p$ system consists of
purely thermal energy, and it is just high enough to approach
the conditions for the onset of multifragmentation.
	On the basis of these data we search for the properties
of the early appearance of break-up events and their competition
with compound-nucleus emission.
	The system $^{56}$Fe+$^{\mathrm{nat}}$Ti is compatible with
an abrasion picture. The excitation energy deposited in the
projectile spectator, still mostly of thermal nature,
establishes the dominance of multifragmentation in the
decay process.
	We will especially discuss the differences in the kinematics
of light-fragment emission in the two systems, conditioned by two
different levels of excitation magnitude.
%
\section{
	Experiment and analysis procedure
}
	The experiment was performed at GSI (Darmstadt).
	A primary beam of $^{56}$Fe was delivered by the heavy-ion
synchrotron SIS at an energy of 1 $A$ GeV.
	The target was constituted of liquid hydrogen
(with a thickness of $87.3$ mg/cm$^2$) contained in a cryostat with
thin titanium windows ($36.3$ mg/cm$^2$ in total), wrapped in thin Mylar foils 
(C$_5$H$_4$O$_2$, total thickness: $8.3$ mg/cm$^2$) for thermal insulation.
	In the target area, other layers of matter intersected the
ion-beam: the accelerator-vacuum window of titanium ($4.5$ mg/cm$^2$) and the
beam-current monitor composed of aluminium foils ($8.9$ mg/cm$^2$).
	In order to disentangle the production and the physical results
related to the interaction with hydrogen from the contribution associated
to the other materials, the whole experimental runs were repeated in 
identical conditions, after replacing the target by titanium foils
having the same thickness of the cryostat windows and wrapped in 
Mylar foils having the same thickness of the cryostat insulation.
	This procedure did not only determine the disturbing 
contributions in the measurement of the $^{56}$Fe$+p$ system, but 
it also provided additional experimental data on an other reaction system.
	With some arbitrariness we name ``titanium target ($^{\mathrm{nat}}$Ti)''
the ensemble of the titanium foils replacing the cryostat window, 
the Mylar wrapping, the accelerator-vacuum window and the beam-current monitor.
	Unfortunately, the measurement of the $^{56}$Fe+$^{\mathrm{nat}}$Ti 
system accounts also for non-titanium nuclei, the pollution of which corresponds 
to their portion in the total number of target nuclei per area and is 
equal to 25.9\%~(Al)~+~7.2\%~(Mylar)~=~33.1\%.  
	It should be remarked that these components are not placed at 
the same distance from the entrance of the spectrometer. 
	Fragments produced in the beam-current monitor or in the 
accelerator-vacuum window could have lower probabilities to be registered 
in the experiment since the angular acceptance is reduced by factors of 
0.33 and 0.25, respectively, compared to products from the titanium 
foils replacing the cryostat.
	Henceforth, we refer to the liquid hydrogen as ``proton target ($p$)''. 
	In this case no polluting contributions are included in the 
final results.
%
\subsection{
	Nuclide identification
}
%
%
\begin{figure}[bb!]
\begin{center}
\includegraphics[width=1\columnwidth]{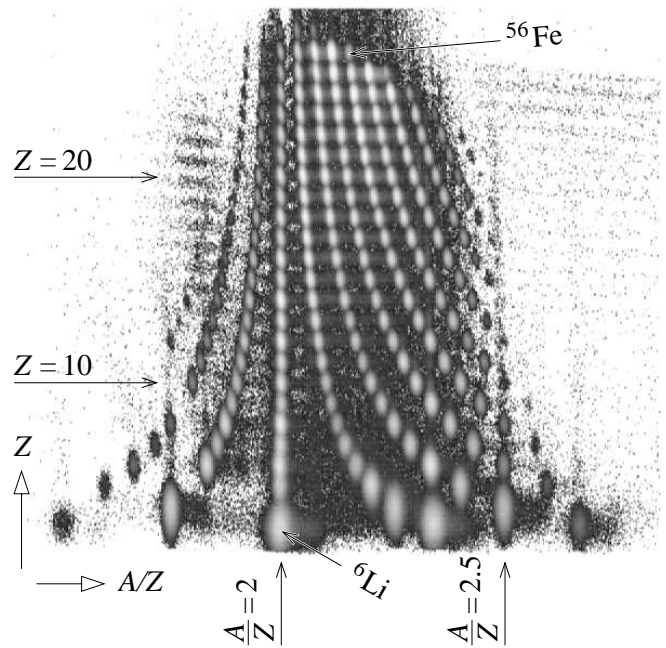}
\end{center}
\caption
{
Experimental isotopic resolution.
The isotopes are grouped in chains. One chain
collects nuclei having the same value of the difference
$N-Z$.
	The straight chain corresponds to $A/Z=2$ or $N-Z=0$.
On the right of this chain the isotopes are neutron rich,
on the left they are proton rich.
}
\label{fig:fig2}
\end{figure}
	The reaction products were analysed with the
high-resolution magnetic spectrometer FRS, constituted of four
dipoles set in an achromatic mode.
	The horizontal positions of the reaction products were deduced
from two plastic scintillators, the
first positioned in a dispersive plane after the first two dipoles,
the second installed in the final achromatic plane.
	From the time of flight ($TOF$), evaluated between the two
scintillators, the relative velocity $\beta$ and the
$\gamma$-factor were obtained.
	The particle charge $q$ was measured with an ionisation chamber
placed in front of the achromatic plane.
	Since the reaction products were fully stripped, the nuclear charge
$Z=q$ was deduced directly.
	The mass $A$ was deduced from the time of flight and the
magnetic rigidity of the particles according to the relation
\begin{equation}
	\frac{A}{Z} = \frac{1}{\mathrm{c}}
		\cdot \frac{\mathrm{e}}{\mathrm{m}_0+\delta m}
		\cdot \frac{B\rho}{\beta\gamma(TOF)}
	\label{eq:equation1}
	\;\; ,
\end{equation}
where $B\rho$ is the magnetic rigidity of a particle, c the
velocity of light, e the elementary charge, m$_0$ the nuclear 
mass unit, $\delta m = \mathrm{d}M/A$ the mass excess per nucleon.
	For the purpose of the isotopic identification, the variation of
$\delta m$ with $A/Z$ can be neglected, and a linear variation of 
$A/Z$ as a function of $B\rho/\beta\gamma$ can be assumed.
	In fig.~\ref{fig:fig2}, the raw data collecting all the
events measured in the experiment are shown.
	Events are ordered according to the measured $Z$ and $A/Z$
so as to obtain an isotopic identification plot.
%
\subsection{
	Longitudinal velocities		\label{subsec:velocities}
}
%
	The measurement of the time of flight is precise enough
for an accurate identification of the mass of the fragments.
Nevertheless, mainly due to the resolution and additionally
due to a slight dependence on the trajectory~\cite{Napolitani01}
it is not suited for a fine measurement of the velocities of the
fragments.
	On the other hand, once an isotope is identified in mass
and charge, a much more precise measurement of the
velocity is obtained directly from the magnetic rigidity of the particle
\begin{equation}
	\beta\gamma = B\rho
		\cdot \frac{1}{\mathrm{c}}
		\cdot \frac{\mathrm{e}}{\mathrm{m}_0+\delta m}
		\cdot \frac{Z}{A}
	\label{eq:equation2}
	\;\;\;\; .
\end{equation}
In this case, the precision of $\beta\gamma$ depends only on
$B\rho$, that has a relative uncertainty of $5 \cdot 10^{-4}$ (FWHM)
for individual reaction products.
	The absolute calibration of the deflection in the magnet in terms of
magnetic rigidity $B\rho$ is performed at the beginning of the
experiment with a dedicated calibration run using the primary beam.
	
\begin{figure}[t!]
\begin{center}
\includegraphics[width=1\columnwidth]{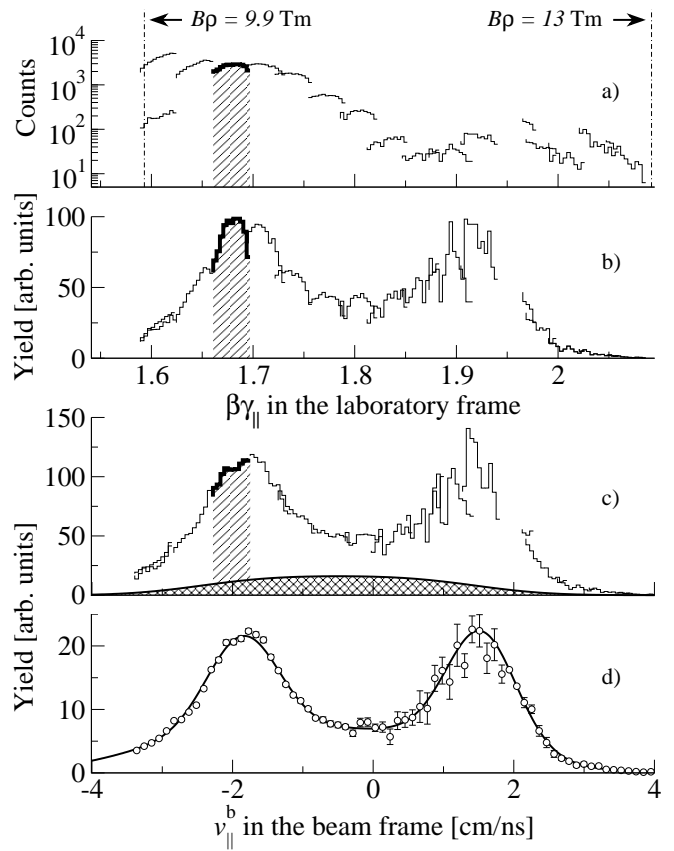}
\end{center}
\caption
{
	Four steps of the analysis procedure to obtain the observed velocity
spectrum of $^{6}$Li emitted in the reaction $^{56}$Fe$+p$.
(a)	Raw spectra of counts as a function of $\beta\gamma$ in the 
laboratory frame. 
	Each segment results from a different scaling of 
the magnetic fields of the FRS.
	One segment associated to the same magnetic scaling
is marked with hatched areas in this plot and in the two following ones.
	Arrows delimit the scanned $\beta\gamma$ range.
(b)	Yields normalized to the same beam dose.
(c)	Elimination of the angular-transmission distortion.
	Spectrum as a function of the longitudinal velocity in the beam 
frame $v^b_{\parallel}$.	
	The broad Gaussian-like hatched area indicates the contributions
from non-hydrogen nuclei.
(d)	All components of the spectrum are composed together averaging 
overlapping points. 
	Contributions from non-hydrogen-nuclei were suppressed.
	The spectrum was divided by the number of nuclei per
area of the liquid-hydrogen target.
	Statistical uncertainties and a fit to the data are shown.
}
\label{fig:fig3}
\end{figure}

\begin{figure}[b!]
\begin{center}
\includegraphics[height=4.2cm]{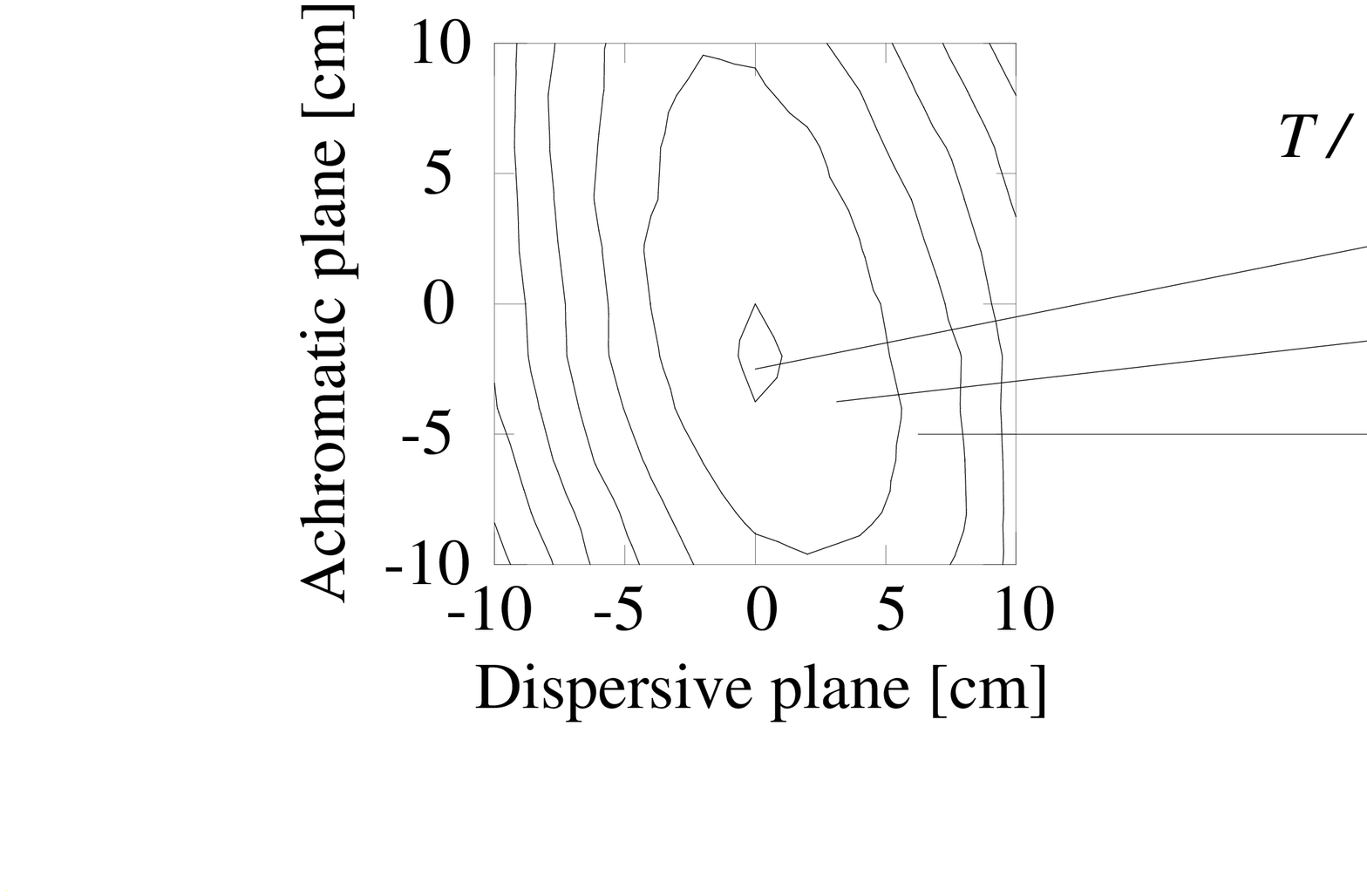}\\
\end{center}
\caption
{
	Transmission of the FRS as a function of the positions
in the dispersive and achromatic planes, relative
to its maximum value. 
	Numerical values are taken from ref.~\cite{Benlliure02}.
}
\label{fig:fig4}
\end{figure}

	Since one single magnetic configuration of the
FRS selects only a $B\rho$ range of about $\pm$ 1.5\%, several
overlapping runs have been repeated imposing different magnetic fields.
	While for the heavy residues close to the projectile
one or few settings were
sufficient to cover the whole velocity spectrum, the light
fragments often required more than ten runs.
	The $B\rho$ scanning of $^{6}$Li, produced in the interaction 
with the target of liquid hydrogen enclosed in the cryostat
constitutes the diagram (a) of fig.~\ref{fig:fig3}:
each segment of the spectrum is obtained from a different
scaling of the set of magnetic fields of the FRS.
	In order to obtain consistent weightings, the counts of the
different measurements were normalized to the same beam dose.
	For each magnetic scaling, this normalization was obtained 
by dividing the corresponding segment of the spectrum by the number 
of projectiles that hit the target during the corresponding run.
	The impinging projectiles were counted with the 
beam-current monitor.
	The renormalized yields are shown in the diagram (b)
of fig.~\ref{fig:fig3}.
%
%
	We should note that the spectrometer accepts only the
fragments emitted in a cone of about $15$ mr around the beam-axis
in the laboratory frame, when the reaction occurs in the 
hydrogen-target position. 
	As a consequence, a light residue like, for example,
$^{6}$Li, generated in a collision at a beam energy
of 1 $A$ GeV can be detected only if emitted with small transverse
momentum.
	The experimental spectrum represents the part of the density
distribution in the velocity space selected by the angular acceptance
of the spectrometer, projected on the longitudinal axis.
	Unfortunately, the angular acceptance depends on the
magnetic rigidity of the particles.	
	As pointed out in the work~\cite{Benlliure02}, for a given set-up
of the spectrometer, the more the intersection of the trajectory of a
particle with the dispersive or the achromatic planes is displaced from the
centers, the lower is the acceptance angle of the FRS.
	The effect appears in the curved sides of each single
segment, with the result of disturbing the overall structure of
the $B\rho$ scanning.
	This distortion, seen in the spectrum of the
plot (b) of fig.~\ref{fig:fig3}, can be successfully corrected by means of
ion-optical calculations that fix the dependence of the angular
transmission on the trajectory.
	The calculation of the ratio of the transmission $T$ relative 
to its maximum value is presented in fig.~\ref{fig:fig4}.
	The corrected spectrum, seen in the plot (c) of
fig.~\ref{fig:fig3}, is the result of scaling up the yields of
the spectrum by the factor $T_{\mathrm max}/T$.
	We also changed from a $\beta\gamma$ spectrum to a
longitudinal-velocity spectrum and, to simplify the analysis, the
reference frame was changed from the laboratory to the beam frame.
	On the average, the projectile interacts in the middle of the
target.
%
%
%
%
\begin{figure}[t!]
\begin{center}
\includegraphics[width=.9\columnwidth]{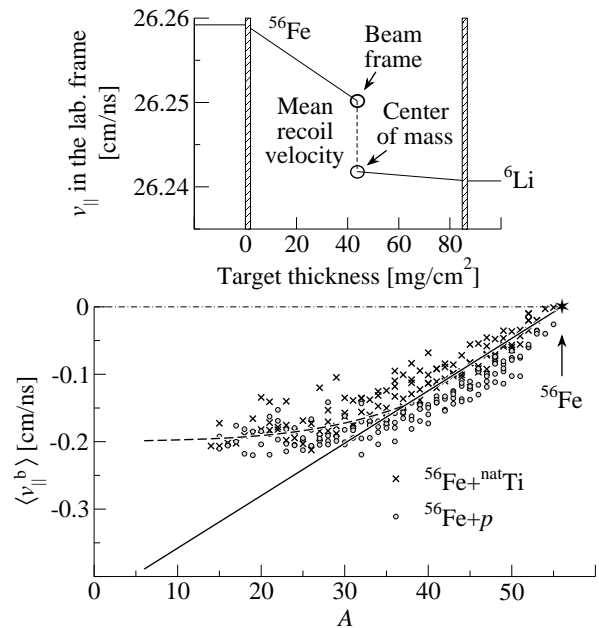}
\end{center}
\caption
{
{\it Top.} Definition of the beam frame and of the center-of-mass
frame of the emitting source with respect to the laboratory
frame.
The diagram corresponds to realistic conditions of the present
experiment for $^{6}$Li.
The solid lines describe the slowing down of the beam
and of the centroid of the velocity spectrum of $^{6}$Li
in traversing the target.
{\it Bottom.} Mean longitudinal recoil velocities in the beam frame
$<v^b_{\parallel}>$ of the reaction residues compared
with the systematics of Morrissey~\cite{Morrissey89} (solid line);
only isotopes with sufficient statistics and entirely measured
velocity spectra are considered.
}
\label{fig:fig5}
\end{figure}
	Therefore, we take into account the slowing down
of $^{56}$Fe in the first half of the target, as represented in
the upper diagram of fig.~\ref{fig:fig5}.
	We also consider that the fragments slowed down in
the remaining half of the target and, therefore, were emitted at
higher velocity than the one we observed.
	The analysis so far illustrated was repeated
for all the isotopes produced in the interaction with the
target of liquid hydrogen enclosed in the cryostat.
	Successively, the same procedure was applied to
the corresponding isotopes produced in the interaction
with the $^{\mathrm{nat}}$Ti target.
	As all spectra are normalized to the same beam dose,
by subtracting the velocity spectra of the residues produced 
in $^{56}$Fe+$^{\mathrm{nat}}$Ti (indicated by the hatched area 
in the plot (c) of fig.~\ref{fig:fig3}) from those
of the corresponding isotopes produced in the target of hydrogen
stored in the cryostat, we could obtain the measured velocity
distributions for the reaction with the liquid hydrogen.
	The resulting yields are unambiguously disentangled 
from any disturbing contributions produced by other material 
present in the target area.
	Finally, the velocity spectra obtained for the $^{56}$Fe$+p$ 
system were divided by the number of nuclei per area of the proton 
target. 
	The resulting spectrum is shown in the diagram (d) of  
fig.~\ref{fig:fig3}.
	In the case of the $^{56}$Fe+$^{\mathrm{nat}}$Ti system,
we should consider that the target is constituted of three
components, the titanium foils replacing the cryostat, the beam-current 
monitor and the accelerator-vacuum window, having 
a number of nuclei per area equal to $n_0$, $n_1$, and $n_2$, respectively.
	We should also recall that these components are placed at
different distances from the entrance of the spectrometer and are
subjected to different values of the angular acceptance, that is about 
$\alpha_0=15.8$ mr, $\alpha_1=9$ mr, and $\alpha_2=7.8$ mr, 
for the layers $n_0$, $n_1$, and $n_2$, respectively.
	Thus, the cross sections given in this work for the "titanium" target 
are calculated using a target composition where the different layers are 
weighted by the corresponding estimated transmission values $T$, assuming 
identical production cross sections in the different target components.
	In particular, the velocity spectra obtained for the
$^{56}$Fe+$^{\mathrm{nat}}$Ti system should be divided by the quantity
$n_0 T(\alpha_0) +  n_1 T(\alpha_1) + n_2 T(\alpha_2)$.

	The experimental data are already complete enough to
let us recognise an important signature of the Coulomb repulsion:
the double-humped spectrum reveals that the velocity of $^{6}$Li
nuclei emitted at small angles has two components:
one appreciably higher and one appreciably lower than the beam.
	According to the references
~\cite{Benlliure01,Enqvist01,Bernas02}, where similar structures
have been observed for fission fragments, we may connect the
double-humped spectrum to the action of the Coulomb field
of a heavy partner in the emission process.

	Once  changed to longitudinal velocities in the
beam-frame $v^b_{||}$, the shift of the
barycenter of the spectrum with respect to zero is equal to the
mean reaction recoil $<v^b_{||}>$.
	Also this quantity, studied in the lower diagram of
fig.~\ref{fig:fig5}, carries a
valuable information about the reaction mechanism, and it can be
related to the friction suffered by the projectile in the
collision, according to a given impact parameter
~\cite{Morrissey89}.
	Due to the limited angular acceptance of the FRS 
which favours the detection of heavier nuclei, a depletion
of the statistics for the measurement
of the lightest nuclei is expected when,
as in this case, the light fragments are
measured together with the heaviest in the same
magnetic setting.
	Due to these problems, the mean velocities
of the light residues can only be determined with
relatively large uncertainties.
	With these large uncertainties, the
information from the mean velocities could not
be exploited.
	Although these mean velocities also enter
into the evaluation of the cross sections,
the uncertainties they introduce are comparable
to those from other sources.
	We preferred to deduce the
mean recoil velocities of lithium, beryllium,
boron and carbon by
extrapolation from the systematics of the data
relative to the ensemble of the heavier residues.
%
\section{
	Results
}
%
\subsection{
	Nuclide cross sections		\label{subsec:v_reconstr}
}
%
%
\begin{figure}[b!]
\includegraphics[width=1\columnwidth]{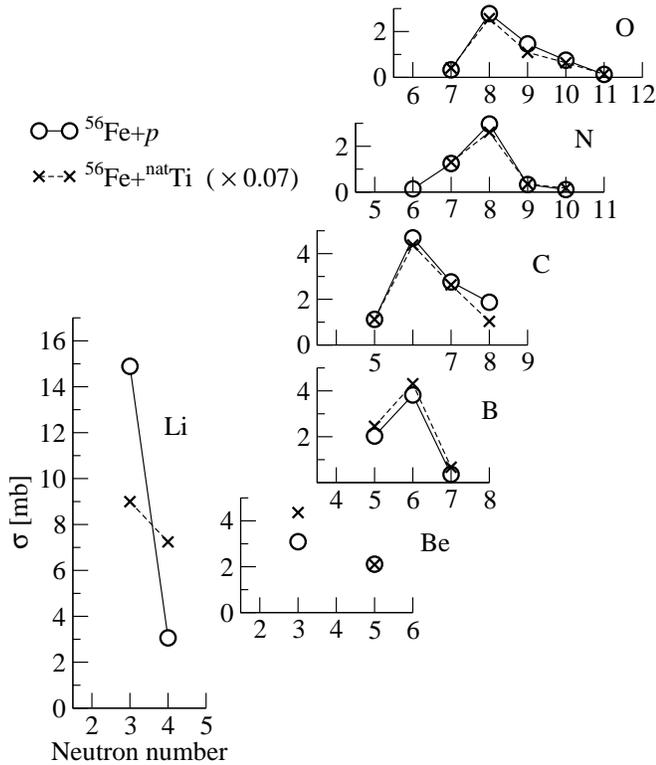}
\caption
{
	Experimental isotopic production cross sections
of some light elements
for the reactions $^{56}$Fe$+p$ and $^{56}$Fe+$^{\mathrm{nat}}$Ti at 1 $A$ GeV.
	The cross sections related to the latter system
are scaled of a factor 0.07.
	Numerical values are collected in table~\ref{tab:tab1}.
}
\label{fig:fig6}
\end{figure}
%
%
%
\begin{figure}[]
\includegraphics[angle=-90, width=\columnwidth]{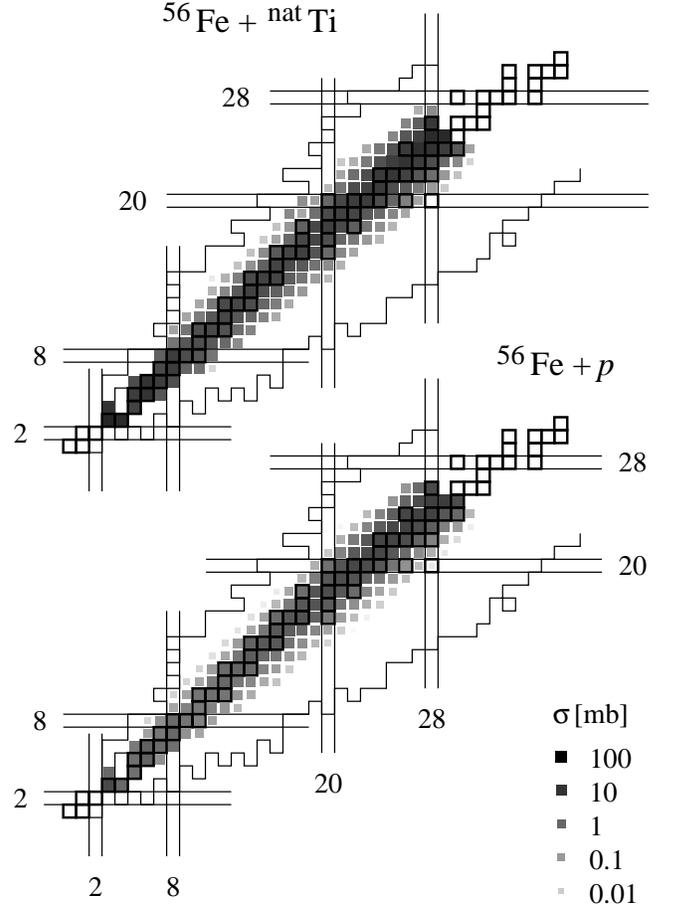}
\caption
{
	Isotopic production cross sections shown on a chart of the nuclides
for the reactions $^{56}$Fe$+p$ and $^{56}$Fe+$^{\mathrm{nat}}$Ti at 1 $A$ GeV.
	The values for $^{54}$Mn in $^{56}$Fe$+p$ and for $^{53}$Cr in 
$^{56}$Fe+$^{\mathrm{nat}}$Ti were obtained from systematics.
}
\label{fig:fig7}
\end{figure}
%
%
\begin{figure}[t!]
\begin{center}
\includegraphics[width=\columnwidth]{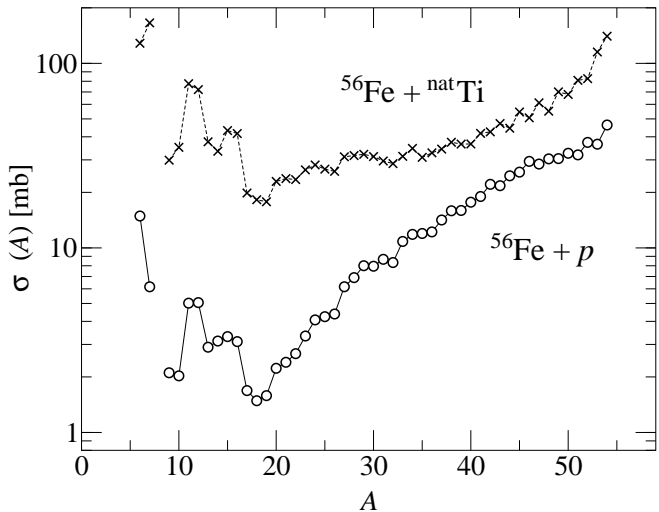}
\end{center}
\caption
{
	Experimental production cross sections
as a function of the mass number.
The statistical uncertainties are lower than 10\%.
The systematic uncertainties evolve from 10\% for the
heavy residues close to the projectile to 20\% for the light
fragments.
}
\label{fig:fig8}
\end{figure}
	When a fragment is emitted with a large absolute velocity
$v=|\vec v|$ in the center of mass, not all the angles of the
corresponding velocity vector $\vec v$ are selected by the finite
angular acceptance of the spectrometer.
	As a result of the data analysis detailed in the previous
section, we obtain the measurement of the apparent cross section
$\mathrm{d}{\mathcal I}(v_{\parallel})/\mathrm{d}v_{\parallel}$
as a function of the longitudinal velocity $v_{\parallel}$.
	This observed cross section differs from the real cross section
due to the angular acceptance.
	The detection of a particle depends on
the perpendicular velocity $v_{\perp}=\sqrt{v^2-v^2_{\parallel}}$
in the center-of-mass frame,
the angle of rotation around the beam direction $\varphi$,
and the velocity $u$ of the center of mass with respect to the
laboratory.
	The dependence on $\varphi$ comes about because the beam pipe
inside the quadrupoles is not cylindrical.
	To reconstruct the full velocity distribution,
independent of the angular acceptance of the spectrometer,
an assumption on the angular distribution is necessary.
	It was concluded from experiments, to which the full angular
range was accessible, that the data are in satisfactory agreement
with an isotropic emission (see, for example, the treatment of
``Moving source analysis'' presented in~\cite{Korteling90}).
	This assumption has been corroborated by a vast collection
of data for reactions of very different nature.
	Isotropic emission has been observed either for lowly
excited fissioning systems~\cite{Moretto89}, or even for very
highly excited nuclei undergoing expansive flow in thermal
multifragmentation~\cite{Karnaukhov99,Karnaukhov03}.
	At least the $^{56}$Fe$+p$ system can be safely included in
this range.
	Slightly less justified is the assumption for
$^{56}$Fe+$^{\mathrm{nat}}$Ti, since some effects of dynamical
multifragmentation could disturb the isotropy.

	Thus, if we assume isotropic particle emission in
the center-of-mass frame,
the variation of the cross section $\sigma(v)$,
as a function of the absolute velocity $v$,
is related to the variation of the apparent cross section 
${\mathcal I (v_{\parallel})}$ as a function of $v_{\parallel}$
by the equation (see appendix~\ref{appendix:v_transmission}):
\begin{equation}
	\frac
	{\mathrm{d}{\mathcal I}(v_{\parallel})}
	{\mathrm{d}v_{\parallel}}=
	\frac{1}{4\pi}
	\int_{0}^{2\pi}
	\mathrm{d}\varphi	
	\int_{|v_{\parallel}|}^
	{\sqrt{v_{\parallel}^2+\widetilde{v}^2_{\perp}(\varphi)}}
	\frac{1}{v}
	\frac{\mathrm{d}\sigma(v)}
	{\mathrm{d}v}\;
 	\mathrm{d}v
	\;\;\;\; ,
	\label{eq:equation3}
\end{equation}
	where $\widetilde{v}_{\perp} = \widetilde{v}_{\perp} (\varphi,u)$
is the highest value of $v_{\perp}$ selected
by the angular acceptance of the spectrometer, and $\varphi$ is the rotation
angle around the beam direction.
	As demonstrated in the appendix~\ref{appendix:v_inversion}, 
it is possible to reverse the relation~(\ref{eq:equation3}) and extract
$\mathrm{d}\sigma(v)/\mathrm{d}v$ by the straightforward solution
of a system of geometric relations.
	The formation cross sections are directly
obtained by integration of $\mathrm{d}\sigma(v)/\mathrm{d}v$.
	In the appendix~\ref{appendix:cross_sections},
	table~\ref{tab:tab1} collects the isotopic cross
sections for the production of light residues,
from lithium up to oxygen, measured in this work for the reaction
$^{56}$Fe$+p$ and $^{56}$Fe+$^{\mathrm{nat}}$Ti.
	The distributions of the formation cross sections evaluated for the
two systems $^{56}$Fe$+p$ and $^{56}$Fe+$^{\mathrm{nat}}$Ti at 1 $A$ GeV are
presented 
in fig.~\ref{fig:fig6} for different elements as a function of
the neutron number and 
in fig.~\ref{fig:fig7}, on the chart of the nuclides.
	The extension of the production appears rather similar
and, in particular, despite the expected difference in
excitation energy reachable in the collisions with the two
different targets, the cross-section distributions of the
residues of the two reactions do not manifest drastic
differences in their features.
	A more quantitative revelation of this similarity is
presented in fig.~\ref{fig:fig8}, where the mass
distributions are compared.
	The difference in the shape of the mass spectra is
significant only for the intermediate masses: the cross section of the
residues of $^{56}$Fe$+p$ decreases from $A=30$ to $A=18$ by about
one order of magnitude, while we observe only a slight decrease by about a
factor of two for
$^{56}$Fe+$^{\mathrm{nat}}$Ti.
	The data reveal that higher excitation energy
introduced by the interaction with titanium, with respect to
proton-induced spallation, results in decreasing the slope of the
mass-spectra in the IMF-range and depleting the cross section for
heavy residues in favour of an enhanced production of light fragments.
	However, the portion of the mass spectra corresponding to
light-particle emission follows a very similar exponential slope
for both systems.
%
\subsection{
	Velocities
}
\begin{figure}[b!]
\includegraphics[angle=0, width=1\columnwidth]{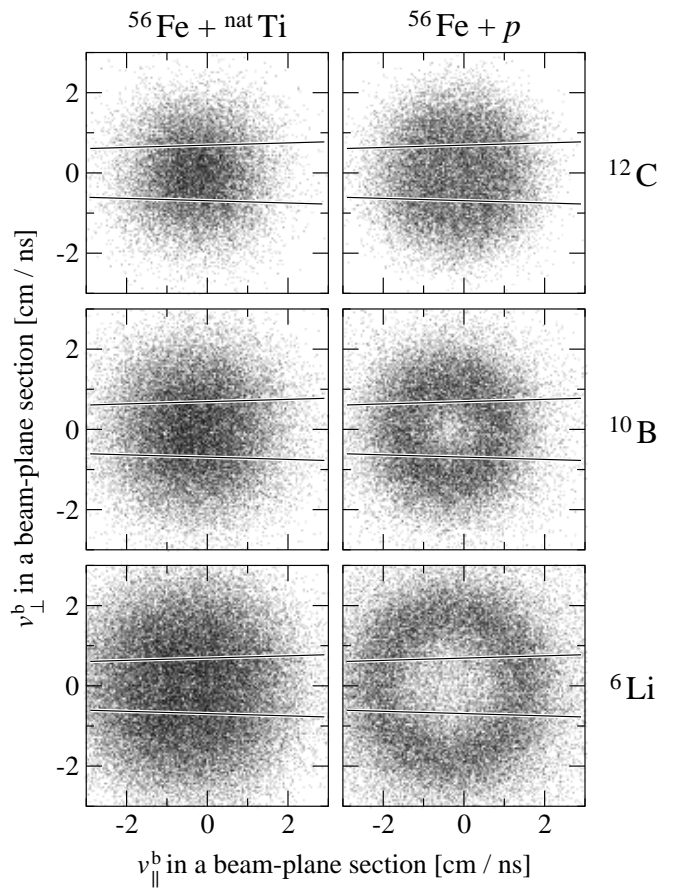}
\caption[\textheight]{
	Reconstructed density plots in velocity space in the beam frame
$(v^b_{\parallel},v^b_{\perp})$ representing the distribution on
a plane containing the beam axis.
	The solid lines denote the angular acceptance
of the spectrometer.
}
\label{fig:fig9}
\end{figure}

%
	In the previous section, the full velocity distributions
were reconstructed from the data and employed to
obtain the residue formation cross sections for the reactions
induced by the proton and titanium target, respectively.
	Though, the cross sections did not yield any
unambiguous distinction between the two reactions that, indeed,
should result into rather different deexcitation pictures on
the basis of the different thermal excitations reached in the two
systems.
	On the contrary, the particularity of the
proton-induced spallation compared to the titanium-induced
fragmentation arises strikingly when the kinematics of the
light-particle emission is investigated.
	The density of velocity vectors $\vec v$ in a plane containing 
the beam axis is presented in fig.~\ref{fig:fig9}.
	(As observed in the following, this presentation is equivalent
to invariant-cross-section plots.)
	Thus, when we compare the $^{56}$Fe$+p$ system to the 	
$^{56}$Fe+$^{\mathrm{nat}}$Ti system on the basis of
the recoil velocity, we find that substantially distinct
mechanisms should be involved in the light-fragment emission in
the two systems.
	In the fragmentation reaction induced by the titanium
target,  all the residues are emitted according to a bell-shape
velocity spectrum.
	A long sequential decay would produce this kind of shape;
in this process, neutrons, protons and clusters are in fact emitted
with different angles with equal probability.
%
%
%
%
%
\begin{figure*}[tp!]
\includegraphics[width=1\textwidth]{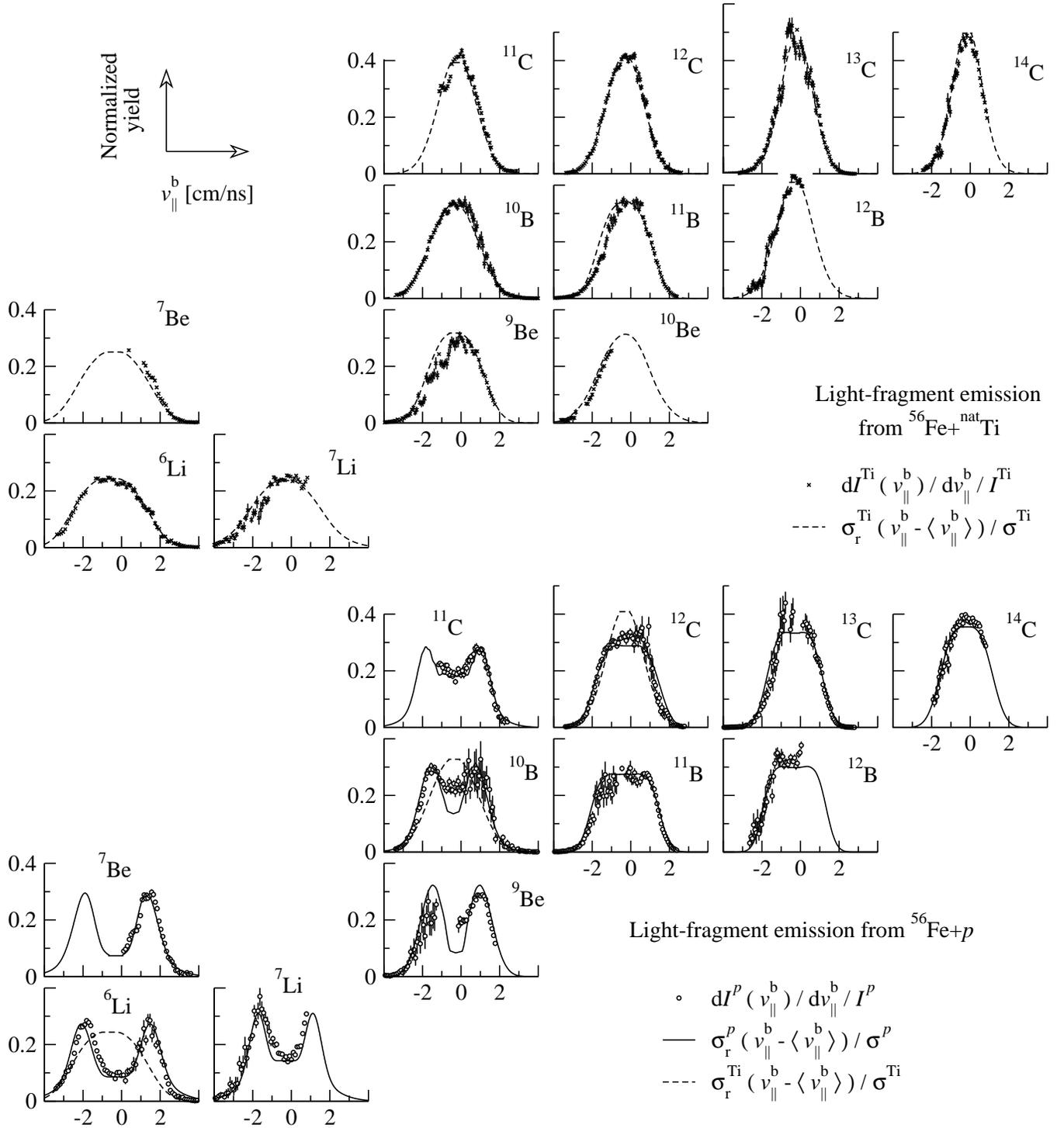}
\caption[\textheight]{
	Velocity spectra of light residues produced in
$^{56}$Fe+$^{\mathrm{nat}}$Ti (upper diagram), 
and in $^{56}$Fe$+p$ at 1 $A$ GeV (lower diagram),
ordered on a nuclear chart.
	They are represented as a function of the velocity in
the beam direction in the beam frame $v^b_{\parallel}$.
	Crosses and points indicate measured spectra
d${\mathcal I}^{\mathrm{Ti}}$/d$v^b_{\parallel}$ and
d${\mathcal I}^{p}$/d$v^b_{\parallel}$, respectively,
defined according to eq.~(\ref{eq:equation3}), and normalized to the unit.
	They represent all fragments transmitted through the FRS.
Reconstructed velocity spectra $\sigma^{\mathrm{Ti}}_r$ and $\sigma^{p}_r$, 
defined according to eq.~(\ref{eq:equation4}) and normalized to the unit 
are marked with dashed and solid lines, respectively.
	In the lower diagram, the reconstructed spectra for 
$^{6}$Li, $^{10}$B and $^{12}$C emitted from $^{56}$Fe+$^{\mathrm{nat}}$Ti 
are superimposed as dashed lines for comparison.
\label{fig:fig10}
}
\end{figure*}
	Nevertheless, due to the high excitation of the hot
fragments gained in such a violent collision,
and due the exponential increase of the cross section of the
light residues with the mass loss, we are in favour of a
multifragmentation picture to depict the dominant
deexcitation process.
	In this case, the hot source is expected to undergo a
fast expansion and successively form several fragments.
	In this scenario~\cite{Bondorf95}, the emission velocity
of a light residue
could vary largely according to different parameters: the
partitioning in the multifragmentation event,
the expansion of the source before the break-up phase, and the
position where the tracked fragment is formed with respect to the
other fragments.
	Also this process would result in a velocity spectrum
with a bell shape centerd at the mean recoil velocity,
equal to the shape we observe.

	On the contrary, when light fragments originate from the
$^{56}$Fe$+p$ system, the reaction dynamics leads to the population of
one most probable emission shell in the velocity space, around the
center of mass.
	This is the case of $^6$Li, as shown in fig.~\ref{fig:fig9}.
	Only a forward and a backward portion of the emission shell
could be measured as selected by the conical cut that the
spectrometer determined: this fully explains the
double-humped velocity spectra shown in
fig.~\ref{fig:fig3}.

	The velocity distributions of the light fragments generated
in the proton-induced reaction carry the unambiguous signature of a
strong Coulomb repulsion in the emission process.
	This observation evidently excludes that the light
fragments could be the final residues of a long sequential
evaporation chain.
	The strong Coulomb component in the emission process
rather reflects the dominating influence of
a very asymmetric split of the source.

	We can now reduce the representation of the recoil-velocity 
distribution $\sigma(v)$ to one dimension,
selecting only those velocities $\vec v$ aligned in the beam
direction, and occupying only abscissae in the plots of
fig.~\ref{fig:fig9}.
	
	Due to our assumption of isotropy, we can define radial
velocity distributions dividing the differential cross section
$\mathrm{d}\sigma(v)/\mathrm{d}v$ associated to a given velocity $v$
in the center of mass by the spherical surface of radius $v$:
%
%
%
\begin{equation}
	\sigma_r(v) =
	\frac{\mathrm{d}^3\sigma}{\mathrm{d}\vec v} =
	\frac{1}{4 \pi v^2}
	\frac{\mathrm{d}\sigma}{\mathrm{d}v}
	\;\;\;\; ,
	\label{eq:equation4}
\end{equation}
	It should be remarked that either in the reference of the center 
of mass or in the projectile frame, $\gamma$ is close to the unit and 
consequently $\sigma_r(v)$ is directly related to the invariant cross 
section $\sigma_I(v)$. 
	Indicating $m = \gamma m$ the mass of the particle, 
$\vec p = \gamma \vec p$ its momentum and $E$ its total energy in the 
center of mass frame (or in the projectile frame), we obtain the equality:
\begin{equation}
	\sigma_r(v) =
	\frac{m^2 \mathrm{c}^2}{m \mathrm{c}^2} 
		\frac{\mathrm{d}^3\sigma}{\mathrm{d}\vec p} =
	\frac{1}{\mathrm{c}^2}E
		\frac{\mathrm{d}^3\sigma}{\mathrm{d}\vec p} =
	\mathrm{c}^{-2} \sigma_I(v)
	\;\;\;\; .
	\label{eq:equation5}
\end{equation}
	Also the planar cuts in velocity space 
$(v^b_{\parallel},v^b_{\perp})$ of fig.~\ref{fig:fig9} are equivalent 
to invariant-cross-section plots~\cite{Babinet81}.

	As a technical remark, the advantage of inverse 
kinematics compared to direct-kinematics experiments 
should be pointed out. 
	The registration of emission velocities close 
to the velocity of the center of mass of the hot remnant 
are not prevented by any energy threshold.
	Thus, only in inverse kinematics we can 
clearly appreciate the gradual transition from a 
chaotic-dominated process, reflected in Gaussian-like
invariant-cross-section spectra, to a 
Coulomb- (or eventually expansion-) dominated process, 
producing a hollow around the center of mass. 
	This characteristic signature we exploit resembles 
the investigation of relative velocity correlations between 
two fragments~\cite{Wang98} in full-acceptance experiments for 
analysing decay times. In that case, the probability to detect 
two almost simultaneously emitted particles in space with small 
differences in direction is suppressed due to the mutual
Coulomb interaction.

	A systematic study of the spectra of
lithium, beryllium, boron and carbon is
presented in fig.~\ref{fig:fig10} and compared
with the observed velocity distributions.
	In the $^{56}$Fe+$^{\mathrm{nat}}$Ti reaction, all spectra show
a bell shape.
	In the $^{56}$Fe$+p$ spallation, the double-humped distribution
appears clearly for isotopes with mass lower than twelve
units.
	The shape of the velocity spectra depends mostly on the mass
rather than on the charge, and chains of isotopes belonging to the
same elements show a transition from a bell shape toward a double-humped
spectrum with decreasing mass.
This transition is not always gradual but, as revealed by the
neighbouring $^{11}$C and $^{12}$C in the lower panel of fig.~\ref{fig:fig10},
sometimes seems to be rather abrupt.
%
\section{
	Discussion			\label{sec:discussion}
}
%
%
\subsection{
	Systematics of kinematical features
}
%
%
%
\begin{figure}[b!]
\includegraphics[width=1\columnwidth]{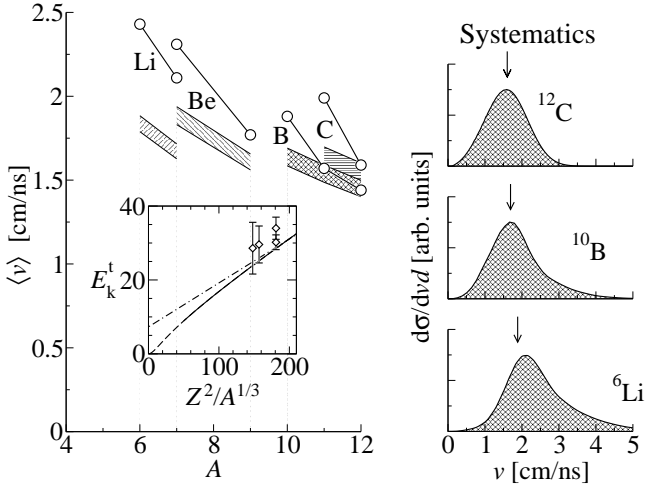}
\caption
{
	{\it Left panel}
	Mean absolute velocities in the reference frame of the
center of mass of the hot remnant, measured for residues of
the $^{56}$Fe$+p$ system (open circles) and deduced from the
systematics of Tavares and Terranova~\cite{Tavares92} (hatched bands).
	The width of the hatched areas results from the
range of the possible mother nucleus from
$^{46}$Ti (lower values) to $^{56}$Fe (higher values).
	In the insert, data points on the total kinetic energy
released in a symmetric split of nuclei close to iron, measured
by Grotowski et al.~\cite{Grotowski84} are compared to the
systematics of Viola~\cite{Viola85} (dot-dashed line) and with the
systematics of Tavares and Terranova~\cite{Tavares92} (solid line).
	{\it Right panel}
	Measured absolute-velocity spectra for the residues
$^{6}$Li, $^{10}$B, and $^{12}$C produced in the $^{56}$Fe$+p$ system.
	The arrows indicate the values obtained by the systematics
of Tavares and Terranova.
}
\label{fig:fig11}
\end{figure}
	On the basis of the ensemble of experimental data
on the production cross section and on the emission
velocity of the residues, we devote this section to
discuss the reaction phenomenology.
	In accordance to the vast literature dedicated to
ion-ion fragmentation (explored in the reviews
\cite{Bondorf95,Richert01}), we can safely relate the
$^{56}$Fe+$^{\mathrm{nat}}$Ti reaction to the formation of highly excited
systems, the decay of which is commonly interpreted as a
multi-body instantaneous disassembly.
	In the following, we will refer to the
$^{56}$Fe+$^{\mathrm{nat}}$Ti collision as a guideline for comparing to a
fragmentation scenario.
	We will rather concentrate on the reaction mechanism of the
proton-induced collision, which points to
competitive types of decay at the same time.

	From a first analysis, the main kinematical characteritics 
of the $^{56}$Fe$+p$ system, recalling a strong Coulomb
repulsion, evocate a binary decay process.
%

%
	On the other hand, other relevant features, like the high
production yields for both light and about half-projectile-mass
residues, could evocate the character of a fast decay, in line with
the scenario of a sudden disassembly of the source depicted for
the $^{56}$Fe+$^{\mathrm{nat}}$Ti system.

	Our first attempt will be to test the pertinence of the
experimental data on the emission velocities with a general
systematics.
	Afterwards, we will discuss the intricacy of the several
possible contributions to the spectral shape of the kinetic-energy
distributions, and the difficulty to extract insight on the
excitation energies involved in the reaction directly from the
measurement.

 \subsubsection{
 	Absolute-velocity spectra
 }
%
	A recurrent analysis of the Coulomb-repulsion aspects is the
comparison of the distribution of absolute velocities of outgoing
fragments $v=|\vec v|$ (where $\vec v$ is the corresponding velocity
vector in the center of mass of the hot remnant) with the systematics
of total kinetic energy released in fission.
	We intend to follow this approach (e.g. \cite{Wang99}) to
test the compatibility of the light-fragment emission in $^{56}$Fe$+p$
with an asymmetric-fission picture.
	The FRagment Separator is particularly efficient in
measuring recoil velocities, because the magnetic rigidity
of the residues is known with high precision
(see section~\ref{subsec:velocities}).
	Indeed, the identity of the mother-nuclei is hidden
in the complexity of the interaction processes related
to high-energy collisions, like the intra-nuclear cascade
and some evaporation events prior to the binary decay.
	The present new data are especially significant as they are
the first measurement of the velocities of fragments
issued of proton-induced splits of iron-like nuclei.
	On the other hand, fission velocities of residues of light
nuclei have been widely investigated in fusion-fission
experiments~\cite{Sanders99}, with the advantage of excluding most
of the ambiguities on the identification of the fissioning nucleus.
	Data on symmetric fission of nuclei close to iron,
formed in fusion reactions were published by Grotowski et
al.~\cite{Grotowski84} and where the basis for the revised
kinetic-energy-release systematics of Viola~\cite{Viola85}.
	This systematics establishes a linear dependence of the most
probable total kinetic energy $E_k^t$ released in a symmetric
fission to the quantity $Z^2/A^{1/3}$, evaluated for the mother
nucleus:
\begin{equation}
	E_k^t = a Z^2/A^{1/3} + b \;\;\mathrm
	\;\;\;\; ,
	\label{eq:equation6}
\end{equation}
where $A$ and $Z$ identify the fissioning nucleus and $a$ and $b$
are parameters fitted to the experimental data
($a=(0.1189\pm 0.0011)$ MeV, $b=(7.3\pm 1.5)$ MeV).
	More recently, new data obtained
for the binary split of even lighter nuclei than iron inspired
Tavares and Terranova~\cite{Tavares92} to revisit the
systematics of Viola once more.
	The new systematics is close to the systematics of Viola
for heavy nuclei down to $Z^2/A^{1/3}\approx 200$.
	As shown in the insert of fig.~\ref{fig:fig11},
iron-like nuclei constitute a turning point: for lower
masses the function changes slope, so that the total kinetic
energy released vanishes for $Z$ approaching 0.
	As anticipated by Viola~\cite{Viola85}, the expectation for
a slope change around iron results by the effect of diffuseness of
light nuclei in disturbing the formation of the neck, in the
liquid-drop picture.
	The following relation was deduced:
\begin{equation}
	E_k^t =
	\frac{Z^2}
	{aA^{1/3}+bA^{-1/3}+cA^{-1}}
	\;\;\;\; ,
	\label{eq:equation7}
\end{equation}
where $a$ $b$ and $c$ are fitting parameters
($a=9.39\mathrm{MeV^{-1}},  b=-58.6\mathrm{MeV^{-1}},  c=226\mathrm{MeV^{-1}}$).
	
	Since the systematics is valid for symmetric splits only, a
term should be added to extrapolate to asymmetric splits, when
two fragments are formed with masses $m_1,m_2$, mass numbers
$A_1,A_2$, and charges $Z_1,Z_2$, respectively.
	Following the hypothesis of non-deformed spheres at contact
(as also imposed in~\cite{Tavares92}), the Coulomb potential is
proportional to the product of the charges of the fission fragments
$Z_1 Z_2$, divided by the distance of their centers, which varies
with $A_1^{-1/3}+A_2^{-1/3}$.
	The conversion from the symmetric to asymmetric
configuration is therefore :
\begin{equation}
	\frac{E_k^t}{E_{k,\mathrm{symm}}^t} =
	\frac{
		Z_1 Z_2 / \left(A_1^{1/3} + A_2^{1/3}\right)
	}{
		\left(\dfrac{Z}{2}\right)^2 \bigg/
		\left[2\left(\dfrac{A}{2}\right)^{1/3}\right]
	}
	\;\;\;\; .
	\label{eq:equation8}
\end{equation}
	It should be remarked that the possible presence of a neck
is not included in this simple relation that, therefore, is a good
approximation for light systems only.
	From the momentum conservation and the introduction of the
reduced mass $\mu = m_1 m_2 / (m_1+m_2)$, we can relate the
total kinetic energy to the velocity $v_1$ of the fragment $A_1$
by the relation $E_k^t = m_1^2 v_1^2 / 2\mu$.
	Introducing the latter form of $E_k^t$ in 
the relations~(\ref{eq:equation8}), and substituting the total kinetic energy
released in symmetric fission with the corresponding value given by
the systematics $E_{\mathrm{syst}}^t$, we obtain the conversion
\begin{equation}
	\frac{v_1^2}{E_{\mathrm{syst}}^t} =
	2^{11/3}\;
	\frac{\mu}{m_1^2}
	\frac{A^{1/3}}{A_1^{1/3}+A_2^{1/3}}
	\frac{Z_1 Z_2}{Z^2}
	\;\;\;\; .
	\label{eq:equation9}
\end{equation}
	Following the strategy of previous publications, e.g.~\cite{Wang99},
for the same light residues we compare the centroids of the measured
absolute velocity spectra to the predicted velocities in fission events;
the latter are deduced from the systematics of total kinetic energy
released in fission by applying the relation~(\ref{eq:equation9}).
	In the right side of fig.~\ref{fig:fig11} we observe
that, while the most probable absolute velocity does not diverge
considerably from the systematics (assumed for $^{56}$Fe as mother
nuclei), the spectra of lighter fragments exhibit a long exponential
tail for very high velocities.
	As a consequence of the asymmetry of the absolute velocity
spectra with respect to a Gaussian distribution, the experimental centroids
lie above the fission systematics, as shown in the left side of
fig.~\ref{fig:fig11}.
	The hatched bands represent the range in velocity due to an
assumed variation of the mother nucleus from $^{46}$Ti (lower
velocities) up to the projectile (higher velocities).
	In previous works (e.g. \cite{Barz86,Wang99,Karnaukhov99})
such tails to very high velocities, reflected in the divergence from
the systematics, were related to the emission from an expanding system
in its initial expansion stage.
%
 \subsubsection{
 	Kinetic-energy spectra
 }
%
%
%
\begin{figure}[b!]
\begin{center}
\includegraphics[width=\columnwidth]{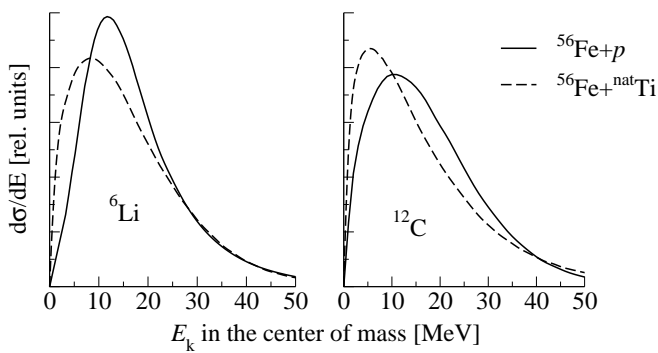}
\caption
{
	Kinetic-energy spectra in the center of mass of the emitting
source obtained from the reconstructed experimental velocity
distributions in the case of emission of $^{6}$Li and $^{12}$C,
respectively.
	The spectra are compared for the $^{56}$Fe$+p$ system (solid lines)
and for the $^{56}$Fe+$^{\mathrm{nat}}$Ti system (dashed lines).
	All spectra are normalized to the same area.
	The smooth distributions result from a spline fit procedure
       to the data.
}
\label{fig:fig12}
\end{center}
\end{figure}
	Directly obtained from the absolute-velocity spectra, the
distributions of kinetic energy $E_k$ offer another representation of
the kinematics, where some more classic features could be searched for.
	In fig.~\ref{fig:fig12} similarities and differences in
kinetic-energy spectra associated with proton and titanium target
nuclei are illustrated.
	The tails to high emission velocities (fig.~\ref{fig:fig11},
right panel) lead to long tails in the kinetic-energy spectra and
characterize both systems.
	We interpret it as a general indication that the collision
generated very high excitation energy in the system.
	It would be tempting to even deduce the thermal properties of 
the system.
	In this case, with particular concern for the $^{56}$Fe$+p$
system, we could draw assumptions on the probability for break-up channels.
	Unfortunately, even if in some studies the nuclear temperature was
deduced from the inverse slope parameter~\cite{Kotov95}, the mixing up
of several effects in the observed kinematics yields serious ambiguities
in the extraction of thermal properties of the source.
	We can list at least eight of the combined effects describing
the observed spectral shapes.
\begin{enumerate}
\item	The presence of a Coulomb barrier results in the deviation of
the spectral shape from a Maxwell-Boltzman distribution (the maximum
moves to higher values).
\item	The transmission through the barrier is ruled by a Fermi
function with an inflection point at the barrier and not by a
discontinuous step function. This effect introduces a widening of the
spectrum.
\item	As a result of the initial stage of the
collision, an ensemble of several possible sources with different $Z$ and $A$
are related to different Coulomb barriers.
	The folding of different Coulomb barrier peaks
results in a broader hump.
\item	If emitted nuclei undergo further evaporation events,
the spectrum widens.
\item	The temperature of the hot source acts on the recoil momenta of
the emitted fragments. If at least major disturbing effects like the
variation of the emitting source, the
Fermi momentum in the hot fragmenting nucleus, described below, 
and the transmission
through the barrier were negligible, it would be possible to deduce the
temperature of the equilibrated fragmenting system from the inverse
slope parameter fitted to the tail of the high side of the energy
spectrum of the residues.
\item	The Fermi momentum of particles removed in the collision with
protons or abraded in the interaction with the titanium target produces
a momentum spread that could be evaluated according to Goldhaber's
formalism~\cite{Goldhaber74}.
\begin{equation}
	\sigma_{p_F}^2 = \sigma_F^2 \frac{A_i(A-A_i)}{A-1}
	\;\;\;\; ,
	\label{eq:equation10}
\end{equation}
where $A$ is the mass of the hot remnant, $A_i$ is the mass of the
emitted cluster and $\sigma_F$ is the Fermi-momentum spread.
	The momentum spread deriving from the Fermi-momentum spread
produces a distribution of momenta of the center of mass of the remnants
in the projectile frame.
	In deducing the energy spectra of the residues in the frame of
the center of mass of the remnant, the spread related to the
Fermi-momentum could not be eliminated as the mass of the remnants are
unknown.
	As a result, the Fermi-momentum contributes both to widening the
spectrum and incrementing the tail for high energies.
	Quoting from Goldhaber~\cite{Goldhaber74}, when a thermalised
system with a temperature $T$ and mass $A$ emits a cluster of mass
$A_i$, the momentum spread of the fragment spectrum is
\begin{equation}
	\sigma_p^2 = \mathrm{m}_0 \mathrm{k} T \frac{A_i(A-A_i)}{A}
	\;\;\;\; ,
	\label{eq:equation11}
\end{equation}
where m$_0$ is the nuclear mass unit and $k$ is Boltzmann's constant.
	The momentum spread $\sigma_{p_F}$ related to the Fermi momentum
adds to the momentum spread induced by the reaction.
	This means that, just reversing the previous relation, the
additional contribution to the temperature related to the
Fermi momentum is equal to the apparent temperature
\begin{equation}
	T_{p_F} = \frac{\sigma_{p_F}^2}{\mathrm{m}_0 \mathrm{k}}\frac{A}{A_i(A-A_i)}
	\;\;\;\; .
	\label{eq:equation12}
\end{equation}
As it was remarked in early studies~\cite{Westfall78},
the extraction of the nuclear temperature from the measured energy
spectra of the residues is therefore a dangerous procedure 
(a recent discussion of the problem of the Fermi motion
is presented in~\cite{Odeh00}).
\item	Multifragmentation events could be accompanied by the expansion
of the nuclear system.
	Nuclei emitted in the initial instant of the
expansion would populate the high-energy tail of the spectrum.
	This is the case for very excited systems~\cite{Siemens79}.
\item	The multiplicity of intermediate-mass fragments simultaneously
emitted might be reflected in the maxima.
	According to previous investigations~\cite{Oeschler00},
a drop in the maximum  energy of the outgoing fragments in a
simultaneous disintegration of the source indicates higher average
multiplicity of intermediate-mass fragments: this is related to the
larger number of participants in the redistribution of the kinetic
energy.
\end{enumerate}
	The last of the enumerated contributions to the energy spectra
is evident in fig.~\ref{fig:fig12}.
	In the proton-induced collision, the position of the maximum
corresponds to larger kinetic energy than in the case of the  titanium
target.
	This might be related to higher multiplicity of
intermediate-mass fragments for the $^{56}$Fe+$^{\mathrm{nat}}$Ti system.
	
	From the analysis of velocity and energy spectra we conclude that 
no clear evidence of the action of a fission barrier could be found.
	Either fission channels are not favoured, or other processes 
obscured them, like additional evaporation stages or the contribution of 
many mother nuclei rather different in mass.
	The most relevant result is the manifestation of high-velocity 
tails, which we interpreted as possible indications of a preequilibrium 
expansion phase.
%
\subsection{
	Nuclear-model calculations
}
%
	We had some hints that very highly excited systems are
formed even in the $^{56}$Fe$+p$ interaction, but we could not
extract quantitative values directly from the experiment.
	We could not recognise the presence of a fission barrier,
but a more complete analysis is required to exclude that solely
compound-nucleus decays are sufficient to explain the light-
fragment production.
	Thus, we wish to carry out a complete reconstruction of the
whole reaction process and compare the ensemble of experimental
results with the calculations.

	Henceforth, we will restrict to the $^{56}$Fe$+p$ system.
	In particular, we will discuss two possible descriptions
for the dominant process of light-fragment formation:
either a series of fission-evaporation decays from a compound
nucleus, or a fast break-up of a diluted highly excited
system, in line with a multifragmentation scenario.
	To fix the initial conditions for the two decay models, we
previously need to calculate mass, charge and excitation-energy
distributions of hot remnants, as these quantities are
not observable in the experiment.
%
\subsubsection{
	Calculation of the excitation energy of the
	hot collision remnants
}
%
%
\begin{figure}[b!]
\begin{center}
\includegraphics[angle=0, width=\columnwidth]{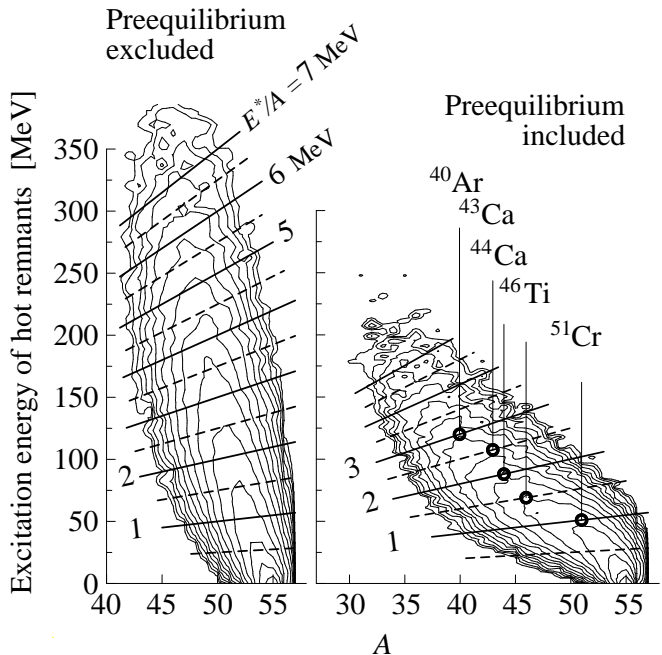}
\end{center}
\caption[\textheight]{
	Hot-fragment distributions generated in the intranuclear
cascade~\cite{Gudima83}, in the case of exclusion (left) and inclusion
(right) of a preequilibrium stage~\cite{Blann71}, for the
$^{56}$Fe$+p$ system.
	Straight lines define constant values of excitation energy
per nucleon, indicated in MeV.
	Five selected isotopes correspond to the most probable mass
and nuclear charge for a given excitation energy per nucleon.
}
\label{fig:fig13}
\end{figure}
%
%
\begin{figure}[t!]
\begin{center}
\includegraphics[width=1.\columnwidth]{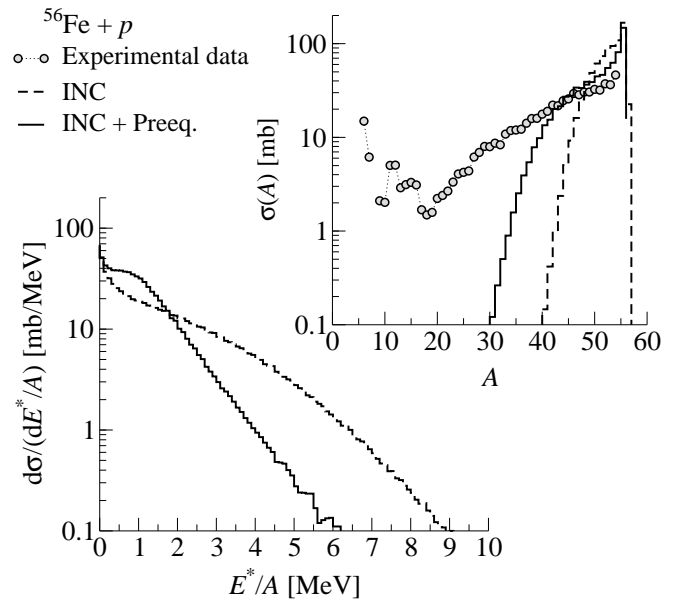}
\end{center}
\caption
{
	Calculated production of the hot fragments after the
intranuclear cascade (modelled according to~\cite{Gudima83})
and a preequilibrium stage (simulated according to~\cite{Blann71}),
for the $^{56}$Fe$+p$ system.
	The cross sections of the hot fragments are shown
as a function of the excitation energy per nucleon $E^*/A$ of the source
(bottom-left) and the mass number (top-right). 
	The mass distribution of cross sections of the hot fragments
is compared to the experimental final-residue production.
}
\label{fig:fig14}
\end{figure}
	The initial non-equilibrium phase of the interaction
$^{56}$Fe$+p$ was described in the framework of
the intranuclear cascade-exciton model developed by
Gudima, Mashnik and Toneev~\cite{Gudima83}.
	The model describes the interaction of an hadron or a
nucleus traversing a heavy ion, considered as a finite open system,
composed of two degenerate Fermi gases of neutrons and protons in a
spherical potential well with diffuse surface.
	The interaction, pictured as a cascade of quasi-free
nucleon-nucleon and pion-nucleon collisions, produces high-energy
ejectiles, that leave the system, and low-energy particles that are
trapped by the nuclear potential.
	As many holes as the number of intranuclear collisions
are produced in the Fermi gas.
	The number of trapped particles and the number of holes
(or excitons, without distinction) determines the excitation energy
of the so-called ``composite nucleus''.

	Hot remnants are often treated as equilibrated or partially
equilibrated systems, both in the case of compound-nucleus
formation and at a freeze-out state.
	Thus, an additional thermalization process might be
necessary to describe the transition from the initial
non-equilibrium phase of the collision to the equilibrium phase
governing the decay.
	Following the hypothesis of the preequilibrium exciton
model, the  intranuclear cascade continues to
develop through the composite nucleus by a sequence of
two-body exciton-exciton interactions, until equilibrium is
attained.
	Two kinds of decay characterize the composite nucleus:
either the transition to a more complicated exciton state, or the
emission of particles into the continuum.
	While in Griffin's model~\cite{Griffin66} all decays are
equiprobable, successive developments proposed more elaborate
descriptions of the competition between decay modes.
	However, when the conditions of the interaction lead to
multifragmentation, the evolution of the composite nucleus is more
complicated, as the system is supposed to expand.
	In the course of the expansion process, an intense
disordered exchange of charge, mass and energy among its
constituents is expected.
	The density of nuclear matter evolves to a more dilute
state, the freeze-out, at which breakup occurs.
	Sophisticated thermal-expansion models were specifically
developed to describe this thermalization
process~\cite{Karnaukhov99,Avdeyev98}.
	Nevertheless, in our calculation we were less specific and
we adopted Blann's preequilibrium exciton model~\cite{Blann71},
independently of the deexcitation scenario.

	In fig.~\ref{fig:fig13} we present a calculation
of the hot-fragment distribution generated in the
intranuclear cascade with and without the inclusion of
a preequilibrium stage, respectively.
	More quantitatively, in fig.~\ref{fig:fig14}
the projections of the distribution are shown as a function
of the excitation-energy-to-mass ratio and mass.
	We observe that preequilibrium is particularly
effective in evacuating part of the excitation energy
and widening the distribution as a function of the mass.
	The hot-fragment mass distribution is compared to the 
measured production of the final residues, in order to
indicate the extension of the deexcitation process.

	When preequilibrium is suppressed, the energy per nucleon
available for the deexcitation largely exceeds 2.5 MeV, a value that
corresponds to the temperature of around 5 MeV, for a fully
thermalised system.
	In this case, the multifragmentation regime is accessible.
	If preequilibrium is included, the average excitation
of the system extends still right up to the expected threshold
for a freeze-out state.
	Hence, according to this previous step of the calculation,
oscillations in direction to break-up decays might be possible.
	Indeed, at these excitation energies break-up channels are
expected to be still in competition with compound-nucleus decay.
	We will proceed to evaluate the extent of this competition
by the use of deexcitation models.
%
\subsubsection{
	Sequential fission-evaporation decay
}
%
%
\begin{figure}[b!]
\begin{center}
\includegraphics[width=1.\columnwidth]{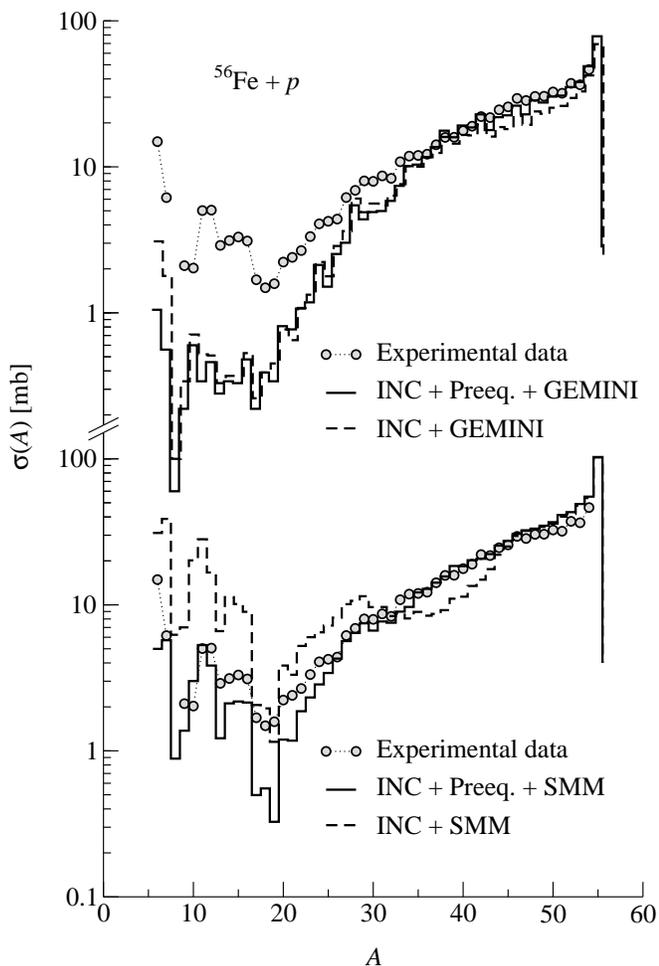}
\caption
{
	Comparison of the measured mass distributions
as a function of the mass number for the system $^{56}$Fe$+p$
with the results of GEMINI (upper part) and SMM (lower part).
SMM is more sensitive than GEMINI to the effect of a
preequilibrium phase.
}
\label{fig:fig15}
\end{center}
\end{figure}
%
%
\begin{figure*}[]
\begin{center}
\includegraphics[width=1\textwidth]{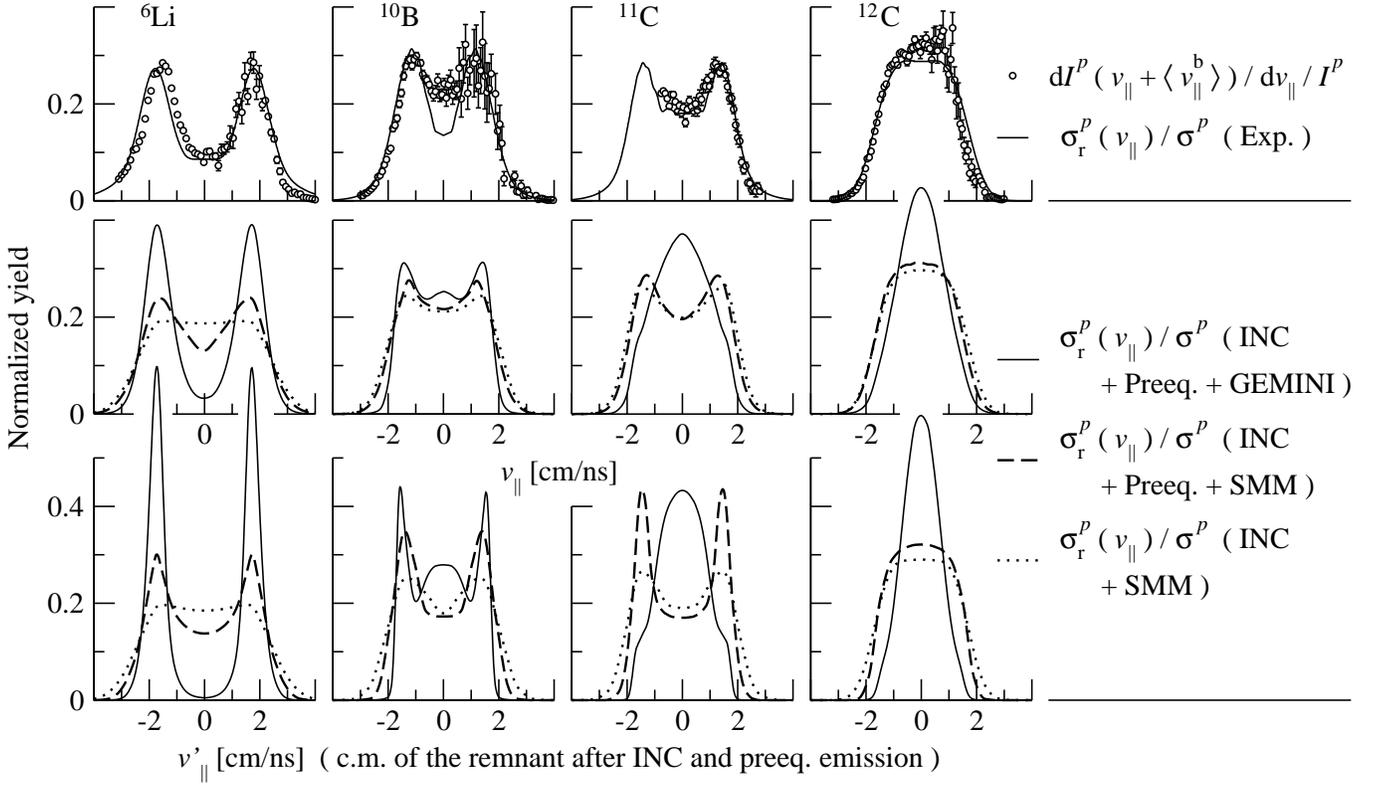}
\caption
{
	{\it First row:} 
	Experimental velocity spectra (circles) and reconstructed
velocity spectra (solid line). 
	Each spectrum is drawn in the reference frame corresponding to 
the measured average velocity value of the fragment considered. 
(This frame corresponds to the "center of mass" frame of the
reaction product drawn in fig.~\ref{fig:fig5}) 
	{\it Second row:} 
	Calculated velocity spectra obtained by GEMINI or SMM following 
INC and the preequilibrium stage, and from SMM following directly INC. 
	Each spectrum is drawn in the reference frame corresponding to 
the calculated average velocity value of the fragment considered. 
	{\it Third row:} 
	Velocity recoil introduced by the GEMINI or SMM phase alone 
(recoils by INC and preequilibrium stages not included). 
	All spectra are normalized to the unit.
}
\label{fig:fig16}
\end{center}
\end{figure*}
	In order to describe the deexcitation process
in the framework of sequential fission-evaporation decays,
we applied the code GEMINI~\cite{Charity88}.
	Within GEMINI a special treatment based on the
Hauser-Feshbach formalism is dedicated to the emission of
the lightest particles, from neutron and proton up to
beryllium isotopes.
	The formation of heavier nuclei than beryllium
is modelled according to the transition-state formalism
developed by Moretto~\cite{Moretto75}.
	All asymmetric divisions of the decaying compound nuclei
are considered in the calculation of the probability of successive
binary-decay configurations.
	The total-kinetic-energy release in fission originally
parameterised according to the systematics of Viola~\cite{Viola85},
eq.~(\ref{eq:equation6}),was replaced by the systematics of Tavares and
Terranova~\cite{Tavares92}, formulated according to the
relation~(\ref{eq:equation7}), and extrapolated for asymmetric
splits by the use of the conversion~(\ref{eq:equation9}).

	We simulated the decay of two possible ensembles
of hot remnants, those issued directly from the stage of
intranuclear cascade, and those which lost part of excitation
energy and mass in a preequilibrium phase.
	The resulting distributions of final residues are
almost indiscernible, revealing that the intermediate-mass
fragments (especially those around oxygen) are not especially
sensitive to the variation of average excitation energy of the
system.
	It might be also pointed out that, when very hot
fragments are allowed to decay by solely fission-evaporation
channels, many nucleons and some light clusters are liberated
at the very beginning of the deexcitation, before eventually
forming an intermediate-mass fragment by fission.
	When the preequilibrium phase is suppressed,
this preliminary emission could constitute a
compensating process.
	In average, the relation between energy loss and
mass loss could be similar in the two processes, and lead to
analogous results.
	The difference is only conceptual, as the
preequilibrium acts on a system still evolving
toward thermalization, and particle evaporation
is connected to a completely thermalised system.
	Only lithium and beryllium revealed a visible
enhancement in the yields with the increase of average
excitation energy.

	The result of the model calculation, compared with
the measured cross sections is presented in the upper side of
fig.~\ref{fig:fig15}.
	The evaluation of the heavy-residue cross sections is
consistent with the experimental data, but a sizeable
underestimation of the production fails to
reproduce the intermediate-mass region.
	Especially the production of the residues populating the
characteristic hollow in the mass distribution reveals to be
generally underestimated by the calculation.
	To complete the comparison, we turn now back
to the first key observable found in our experimental investigation:
the velocity spectra of light fragments.
	In the first row of fig.~\ref{fig:fig16} the experimental
spectra of $^{6}$Li, $^{10}$B, $^{11}$C and $^{12}$C are shown,
together with their velocity reconstruction (solid line).
	Within GEMINI, all decays are decorrelated in time and when
more fragments are produced they do not interact in the same Coulomb
field.
	Binary compound-nucleus emission is connected with a restricted
range of heavy sources close to the projectile mass, reflected in the
small width of the Coulomb peaks, as shown in the second row of
fig.~\ref{fig:fig16}.
	This feature characterizes only the formation of the lightest
fragments and disappears with increasing mass of the residues.
	The calculations presented in the second row of
fig.~\ref{fig:fig16} should not be compared to the experimental data.
	The effect of the Coulomb repulsion involved in the deexcitation
and disentangled from the smearing effect of the intra-nuclear cascade and
preequilibrium emission can be appreciated in the third row of
fig.~\ref{fig:fig16}, where the reference frame has been fixed to the 
center of mass of the initial system formed at the beginning of the 
fission-evaporation process. 
	In the calculation, the transformation of the two Coulomb peaks
into one single wide hump occurs for lower masses than experimentally
observed.
	The model generates one single hump in the longitudinal velocity
spectra of light fragments when a longer evaporation cascade is
involved, and characterized by mainly alpha and nucleon emission.
	Moreover, the total width of the calculated spectra is narrower
than observed.
\subsubsection{
	Fast break-up
}
%
	We imputed the underestimation of intermediate-mass
fragment formation to an incomplete description of the most
highly excited decaying systems when solely
fission-evaporation deexcitation was considered.
	In this respect, we turned to the Copenhagen-Moscow 
statistical multifragmentation model (SMM)~\cite{Botvina85,Bondorf95},
that is the extension of the
standard statistical evaporation-fission picture toward high
excitation energies, treated by adding the fast simultaneous
disassembly of the system as a possible decay channel.
	The hybrid model of intranuclear cascade followed by SMM
was already applied in previous studies of proton-induced 
reactions~\cite{Botvina85bis,Botvina90} for the description 
of similar experimental data.
	In the framework of SMM, the evaporation from the
compound and compound-like nuclei is included and, therefore,
at low excitation energies, if the channels with production of
compound-like nuclei dominate, SMM gives results similar to
GEMINI.
	In particular, the statistical cluster evaporation
is treated within the Weisskopf formalism, extended to the
emission of nuclei (in their ground state or available excited
states) up to $^{18}$O~\cite{Botvina87}.	
	On the other hand, when very high excitation
energies are reached in the collision, the system is assumed to
be diluted and to have attained the freeze-out density $\rho_b$.
	In previous studies~\cite{Bondorf85} $\rho_b$ was 
calculated to evolve as a function of the excitation energy per 
nucleon toward an almost asymptotic value equal to $1/3$ of the 
ground-state density $\rho_0$ for high excitation energies 
($E^*/A$ > 5 MeV).
	In the present calculation, an energy-dependent free 
volume is used to determine the probability for different break-up 
partitions. 
	On the other hand, for the calculation of the Coulomb
interaction among fragments the freeze-out density $\rho_b$
is introduced as a fixed quantity, equal to the asymptotic value
$\rho_b =\rho_0/3$.
	According to the physical picture, when the region of
phase (spinodal) instability is reached, at least partial
thermodynamic equilibrium is expected and the fragment formation
takes place according to chaotic oscillations among different
break-up configurations, from event to event.
	In SMM, within the total accessible phase space, a
microcanonical ensemble of all break-up configurations,
composed of nucleons and excited intermediate-mass fragments
governs the disassembly of the hot remnant.
	The probability of different channels is proportional
to their statistical weight.
	Several different break-up partitions of the system
are possible.

	In fig.~\ref{fig:fig15}, the calculation based on SMM reveals
to better describe the reaction in comparison to GEMINI.
	It should also be observed that the production of
intermediate-mass fragments is sensitive to the excitation energy
of the source.
%
%
\begin{figure}[pt!]
\begin{center}
\includegraphics[angle=0, width=\columnwidth]{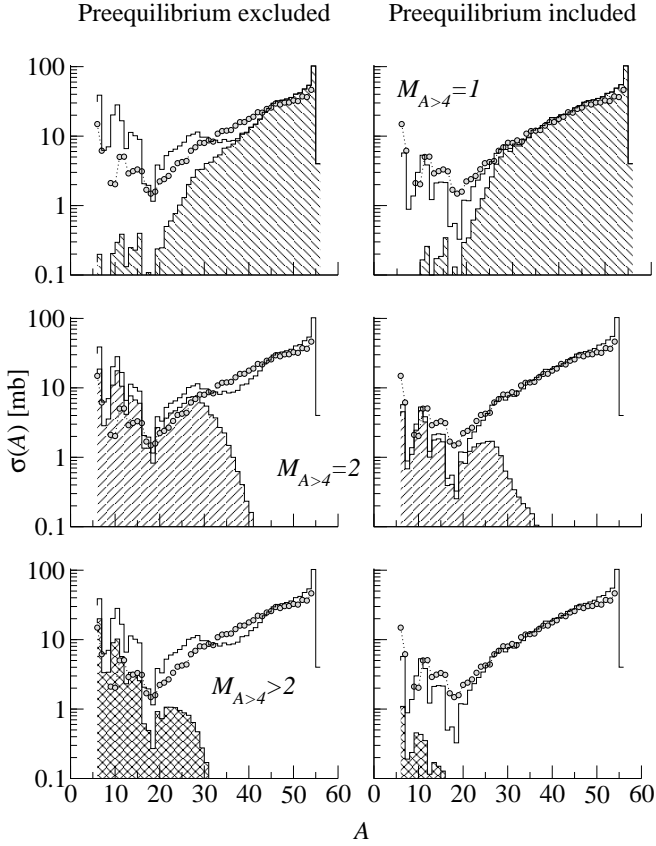}
\caption[\textheight]{
	The hatched areas represent portions of the residue
production calculated with SMM, subdivided according to different
multiplicities of intermediate-mass fragments (having $A$>4).
	The total production measured experimentally
(dots) and calculated (solid line) is superimposed for comparison.
	The calculation disregards preequilibrium in the left
diagrams and includes preequilibrium in the right diagrams.
}
\label{fig:fig17}
\end{center}
\end{figure}
%
%
\begin{figure}[b!]
\begin{center}
\includegraphics[angle=0, width=1\columnwidth]{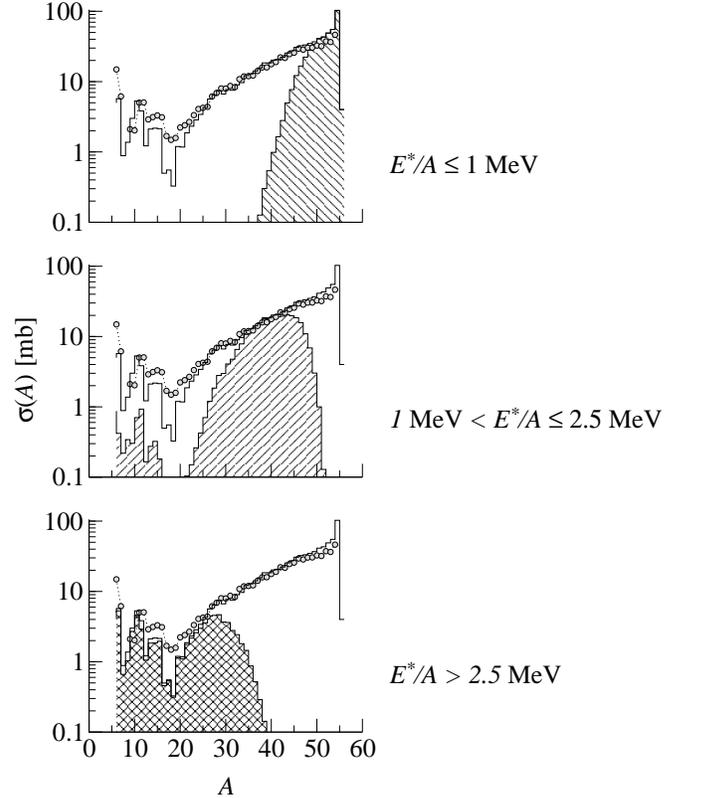}
\caption[\textheight]{
	Different portions (hatched areas) of the residue
production calculated with SMM are selected according to
different ranges in the excitation energy per nucleon $E^*/A$
of the source.
	The calculation is performed only for the case of
inclusion of preequilibrium.
	The total production measured experimentally
(dots) and calculated (solid line) is superimposed for comparison.
}
\label{fig:fig18}
\end{center}
\end{figure}
	A more detailed view on the reconstruction of the
reaction mechanism is presented in fig.~\ref{fig:fig17}, where
the multiplicities involved in the fragment formation are
investigated.
	The major cross sections are fully determined by
evaporation decays.
	This is true for the calculation where preequilibrium
is included.
	On the contrary, when preequilibrium is excluded,
a depletion  of the heavy evaporation residues arises as a
result of the excessive enhancement of higher-multiplicity modes
(cluster emission and multifragmentation).
	The intermediate-mass fragments are almost totally
produced in binary decays of mainly break-up character.
	Multifragment emission channels (with multiplicity mainly
equal to three) have a minor contribution in the decay of the
thermalised system but, when preequilibrium is disregarded,
their incidence is in strong competition with binary splits.
	We might conclude that a consistent description of
the production of fragments could be found in between
these two approaches.

	We can extend the investigation to the excitation
energies connected to the production of fragments with different masses.
	According to the calculation presented in fig.~\ref{fig:fig18}
(now performed only including the preequilibrium phase),
intermediate-mass fragments are almost all formed in the decay of
highly excited remnants, with excitation energy per nucleon
above $2.5$ MeV.

%
%
\begin{figure}[tb!]
\begin{center}
\includegraphics[angle=0, width=1\columnwidth]{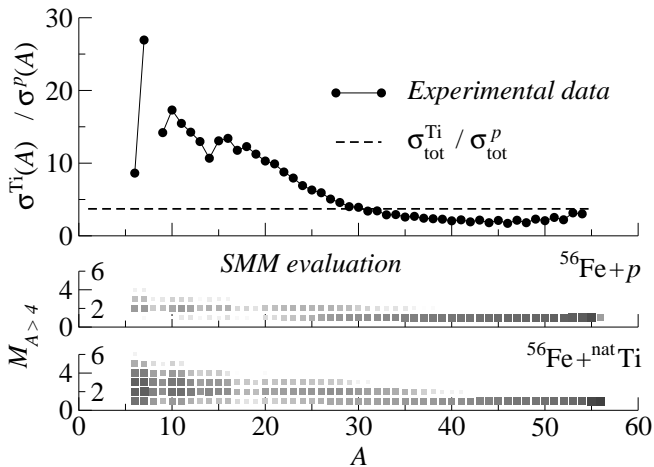}
\caption[\textheight]{
	{\it Upper part.}
	Experimental data on the production cross-sections 
ratio as a function of the mass number for the
reaction $^{56}$Fe+$^{\mathrm{nat}}$Ti versus the reaction $^{56}$Fe$+p$.
	The data are compared with the ratio of total nuclear
cross sections for the two reactions, calculated according
to the model of Karol~\cite{Karol75}.
	{\it Lower part.}
	SMM calculation of the probability for the formation
of a residue as a function of the mass number and the multiplicity.
	The multiplicity is intended as the number of
projectile-like residues heavier than an alpha particle produced
in one collision.
}
\label{fig:fig19}
\end{center}
\end{figure}

	An experimental indication of how multiplicity is related
to the excitation energy is suggested by
fig.~\ref{fig:fig8} and the upper diagram of fig.~\ref{fig:fig19},
where the ratio of the sum of the individual
production cross section in the two reactions
and the total cross-section ratio are compared.
	The latter is calculated according to the model of
Karol~\cite{Karol75}.
	The production cross-section ratio scales with the total
cross-section ratio only in the region of higher masses, presumably
coming from more peripheral collisions, while it deviates for
lighter masses.
	The deviation must be related to the different mean
multiplicities in the two reactions: lighter masses are more populated
in the $^{56}$Fe+$^{\mathrm{nat}}$Ti reaction, and great part of this
increase might be related to higher multiplicities.
	The observation of the gradual increase of
multiplicity with the excitation of the system, verified in the
calculation of different yield spectra associated to different
energy ranges of hot remnants, could be followed further when
extending to the $^{56}$Fe+$^{\mathrm{nat}}$Ti reaction.
	As pictured in the lower plots of fig.~\ref{fig:fig19},
this behaviour is well reproduced by calculations with SMM.
	In this respect, the light-residue production characterizing
the $^{56}$Fe$+p$ system might just be interpreted as the early onset
of the process that will govern the decay of the
$^{56}$Fe+$^{\mathrm{nat}}$Ti system.

	To conclude this section on model calculations, we focus 
once more on the velocity spectra of the light fragments shown 
in fig.~\ref{fig:fig16}.
	We already discussed the difficulty to combine the
wide shapes of the velocity spectra and their mean values with
a fission barrier.
	We inferred that the extensions of the velocity distributions
to very high velocities might reflect higher kinetic energies than
an asymmetric fission process could release.
	Consistently with this expectation, on the basis of model
calculations we could connect the production of light residues to
very high excitation energies of the source.
	Above around $2.5$ MeV of excitation energy
per nucleon, the process of light-residue
production is still presumably dominated by binary decays, but the
contribution of disintegration in more fragments is not excluded.
	In this case, parts of the distribution corresponding
to smaller velocities should be more populated than in a purely
binary split.
	In the representation of fig.~\ref{fig:fig16}, this
contribution would fill more central parts of the spectra
when lower asymmetry characterizes the break-up partition.
	In a statistics of events where three about-equal-size
fragments are produced simultaneously, the velocity spectrum of any
of them will be Gaussian-like.
	If three fragments are produced, of which two are
considerably lighter than their heavy partner, the velocity
spectra $\sigma_r$ of the two light ejectiles will be double humped.
	Another contribution in populating lower velocities
could be associated to different break-up configurations, where
the partner or the partners of the light residue have
different masses.
	In this case, the spectrum is the folding of several
binary-like components characterized by different spacing
between the two maxima and different widths around the maxima,
all this resulting in the superposition of two (backward and forward)
triangular-like distributions that could eventually merge in
a general bell-shape.
	We might also consider standard evaporation cooling down
the break-up residues, emitted in some excited states.
	In this case, the secondary ``slow'' emission process operates
outside of the common Coulomb field of the fragmenting remnant,
and would produce a general widening of the spectrum around
its maxima.
	As portrayed in the second row of fig.~\ref{fig:fig16},
SMM describes very consistently the experimental spectra.
	In the third row the effect of the Coulomb interaction
and eventually the expansion is illustrated by referring to 
the center of mass of the system formed right after the intra-nuclear
cascade and the preequilibrium, if included.
	SMM calculates the Coulomb interaction between 
fragments by placing them inside the freeze-out volume $\rho_b$.
	Contrarily to GEMINI, it takes into account different positions 
of the fragments, including two-body and many-body partitions.
	Some multifragmentation channels may resemble a two-body 
process even at relatively high excitation energy.
	These channels can also include additional
small fragments which may look like evaporation ones. 
	However, these additional small fragments can essentially 
change the Coulomb interaction in the volume
and the thermal energy in the system, and influence the
kinetic energies of the main two fragments. 
	The binary character characterizing the experimental results
is properly reproduced and the velocity distributions calculated with 
SMM are wider than those obtained from GEMINI.
%
%
\begin{figure}[h!]
\includegraphics[width=.8\columnwidth]{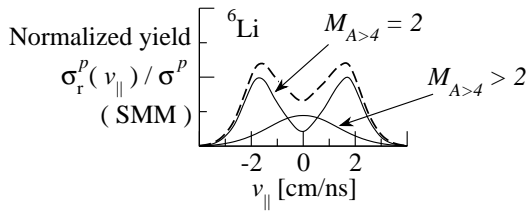}
\caption
{
	Contribution of different multiplicity channels (A>4)
	to the velocity spectrum of $^6$Li, as calculated by SMM.
	The representation is the same as in the second column of 
	fig.~\ref{fig:fig16}.
}
\label{fig:fig20}
\end{figure}
	As shown in fig.~\ref{fig:fig20}, the gradual 
filling of the center of the spectra could be related to different 
break-up configurations and to possible multibody disintegration.
	The abrupt change of shape in passing from $^{11}$C to $^{12}$C
(correctly reproduced by SMM) might be related to the more favoured
evaporation channel toward the formation of $^{12}$C, that
could collect several different decay processes and evaporation
decays from neighbouring nuclei.
	This preferential decay toward $^{12}$C smears out
any binary character of the spectrum.	
	As the break-up configuration varies with the mass and the charge
of the end-product, it varies also with the excitation energy available
for the disassembly of the hot remnant.
	This was evident in fig.~\ref{fig:fig17}, where we compared the 
production cross sections of the residues, and it is also evident
in fig.~\ref{fig:fig16}, by analysing the velocity spectra.
	It is evident that the suppression of the preequilibrium
induces a smearing effect on the spectra. 
	This effect is dramatic for $^{6}$Li as the double-humped
spectrum is completly smeared out in one large single hump with a
flat top, and it becomes similar to the velocity spectrum of $^{6}$Li
produced in the even more excited $^{56}$Fe+ $^{\mathrm{nat}}$Ti system.
%
\section{
	Conclusion
}
%
	The mechanisms of light-particle emission from two systems,
$^{56}$Fe$+p$ and $^{56}$Fe+ $^{\mathrm{nat}}$Ti at 1 $A$ GeV, have been investigated.
	The latter was regarded as a baseline for very high-energy
processes (multifragmentation).
	We focused mainly on the proton-induced reaction.	
 	The certain understanding that we could attain from the
analysis of the $^{56}$Fe$+p$ system is that light residues are produced in
the decay of highly excited remnants.
	Furthermore, from a more quantitative discussion, we inferred
that the emitting source should also be heavy and close to the projectile
mass.
	The magnitude of the Coulomb repulsion, together with the
very high formation yields even suggested that an asymmetric
break-up process, hardly connected to asymmetric fission or statistical
cluster emission, might be the favoured channel of light-residue production.
	These findings were derived from experimental observables like the
isotopic cross sections measured for the whole ensemble of the residues,
and the velocity distributions of the emitted fragments
in the projectile frame along the beam axis.
	Especially the shape of the velocity spectra offered us a
microscopic insight into the mechanisms of light-particle emission.
	In analysing the features of the velocity spectra,
we failed in describing the kinematics within a
general systematics of fission total-kinetic-energy release.
	A complete simulation of the whole reaction process,
where sequential fission-evaporation decays govern the deexcitation,
could not consistently describe the gross experimental features of the
decay.

	We suggested that the characteristics of the
kinematics and the production of light particles could carry
indications of fast asymmetric splits.
	A novel description of the complete reaction process,
including channels of fast break-up decays revealed to be
more adapted in depicting the decay of the most highly excited
remnants, and was compatible with the high yields for light residues
and the complex shapes of the velocity spectra.
	Encouraged by this consistency and, first of all,
on the basis of previous theoretical and experimental
results (see references in the section~\ref{sec:discussion}),
we suggested that protons at incident energies of 1 $A$ GeV
traversing iron-like nuclei can introduce very high thermal excitation
energy per nucleon in the system, even above 2.5 MeV.
	Such a thermal excitation could lead to attain
freeze-out conditions.

	We assume that if the excitation energy is high but not
sufficient to lead to freeze-out, the system might still expand, but
without reaching the break-up phase, the expansion subsequently
reverses to compression in a path toward the formation of a 
compound nucleus .
	This behaviour might occur before the nucleus cools
down by fission-evaporation decays.
	On the contrary, when the excitation energy is just sufficient
to access break-up channels, partitions with low multiplicity of
intermediate-mass fragments and high asymmetry are favoured:
the decay results mainly in the simultaneous formation of one heavy
residue, with mass close to the hot remnant, and one or more
light clusters and nucleons.
	As an extreme case, two fragments rather asymmetric in mass may
be formed in the same fast break-up process.
	The formation of light fragments in the $^{56}$Fe$+p$ reaction
could be explained by this picture.
%
\section{
	Acknowledgments
}
%
	The experiment was performed in a joined effort by an international
collaboration to provide new data for nuclear technology and for
astrophysics. 
	We thank our colleagues
L. Audouin, C.-O. Bacri, J. Benlliure, B. Berthier, A. Boudard, 
E. Casarejos, J.J. Connell, S. Czajkowski, J.-E. Ducret, T. Enqvist, 
T. Faestermann, B. Fernandez, L. Ferrant, J.S. George, F. Hammache, A. Heinz,
K. Helariutta, A.R. Junghans, B. Jurado, D. Karamanis, S. Leray, 
R.A. Mewaldt, M. Fern\'andez Ord\'o\~nez, J. Pereira-Conca, M.V. Ricciardi, 
K. S\"ummerer, C. St\'ephan, C. Villagrasa, F. Viv\`es, C. Volant, 
M.E. Wiedenbeck, and N.E. Yanasak.
	
	We are especially indebted to M. Bernas and P. Armbruster
for sharing with us their expertise in the study of reaction
kinematics and for enlightening explanations on the ion-optics of
the FRagment Separator.
	We are grateful to A. Keli\'c and O. Yordanov for precious 
suggestions and remarks about the data analysis.	
	The interpretation of our results profited from stimulating
discussions with A. Boudard, C. Volant and S. Leray.
	We wish to thank F. Gulminelli for carefully reading the 
manuscript as well as B. Borderie and J. P. Wieleczko for fruitful 
discussions.
	This work was supported by the European Union under the 
programme "Access to Research Infrastructure action of the Improving 
Human Potential" contract EC-HPRI-CT-1999-00001.
%
%
%
\appendix
%
\section{
	The velocity reconstruction	\label{appendix:v_reconstr}
}
%
\subsection{
	Demonstration of the relation~(\ref{eq:equation3})
				     \label{appendix:v_transmission}	
}
%
\def\vpar{v_{\|}}
\def\vper{v_{\bot}}
\def\vperm{\widetilde{v}_{\bot}}
\def\diff{\mathrm{d}}
\def\ubound#1{\sqrt{#1^2+\vper^2}}
\def\dsdv{\displaystyle{\diff\sigma\over\diff v}}
\def\dsdvdo{\displaystyle{\diff^3\sigma\over\diff v\,\diff\Omega}}
\def\dsdvdol{\displaystyle{\diff^3\sigma/(\diff v\,\diff\Omega)}}
\def\dyield{\diff{\cal I}(v_{\|})}
\def\yield{\displaystyle{\dyield\over\diff\vpar}}
\def\ud{\mathrm{d}}
	We describe hereafter the derivation of the equation~(\ref{eq:equation3}), 
which connects the measured spectra $\dyield/\diff\vpar$ as a function of the 
longitudinal velocity component $\vpar$ in the center-of-mass frame to the 
cross-section variation in velocity space in the center-of-mass frame.
	In a general case, the latter distribution is not isotropic,
but a function of the absolute velocity $v$, the polar angle from the 
beam direction $\theta$, and the azimuthal angle around the beam axis
$\varphi$. It will be denoted as $\dsdvdol$, where $\Omega$ is the solid angle.
	The velocity component orthogonal to the beam axis is $\vper$.
	The contribution to the experimental yield in the interval
$[\vpar,\vpar+\Delta\vpar]$ is obtained by integrating $\vper$ in the slab
orthogonal to the beam axis~:
\begin{eqnarray}
	\yield	&=&\iint\frac{\diff^3\sigma}{\diff\vec v}
		\,\vper\,\diff\vper\,\diff\varphi\\
		&=&\iint\frac{1}{v^2}\dsdvdo
		\,\vper\,\diff\vper\,\diff\varphi
	\;\;\;\; .
	\label{eq:equation13}
\end{eqnarray}
	For the orthogonal velocity integration the lower limit is 0 and the 
higher limit is related to the angular acceptance of the spectrometer. 
	Since the latter is not necessarily circular, it can depend on 
$\varphi$ and will be denoted as $\alpha(\varphi)$. 
	The maximal orthogonal velocity may be derived from the Lorentz 
transformation of the momentum and it 
reads~: $\vperm(\varphi)=\gamma(u+\vpar)\alpha(\varphi)$, where $u$ and 
$\gamma$ are the velocity and the Lorentz factor of the center of mass in the 
laboratory frame, respectively.
	Introducing these limits in the integration, we write~:
\begin{equation}
	\yield = \int\limits_{0}^{2\pi}\left[
	\int\limits_{0}^{\vperm(\varphi)}{1\over v^2}\,\dsdvdo\,
	\vper\,\diff\vper\right]
	\diff\varphi
	\;\;\;\; .
	\label{eq:equation14}
\end{equation}
	Changing the integration variable from $\vper$ to 
$v=\sqrt{\vpar^2+\vper^2}$ we obtain~:
\begin{equation}
	\yield=\int\limits_{0}^{2\pi}\left[
	\int\limits_{\vert\vpar\vert}^{\sqrt{\vpar^2+\vperm^2(\varphi)}}{1\over v}\,
	\dsdvdo\,\diff v\right]\diff\varphi
	\;\;\;\; .
	\label{eq:equation15}
\end{equation}
	In the case of an isotropic velocity distribution, $\dsdvdol$ reduces to 
$(1/4\pi)(\diff\sigma/\diff v)$ which, substituted in eq.~(\ref{eq:equation15})
gives eq.~(\ref{eq:equation3}).
%
\subsection{
	Reversibility of the relation~(\ref{eq:equation3})	
					\label{appendix:v_inversion}	
}
%
\def\vpar{v_{\|}}
\def\vper{v_{\bot}}
\def\vperm{\widetilde{v}_{\bot}}
\def\diff{\mathrm{d}}
\def\ubound#1{\sqrt{#1^2+\vper^2}}
\def\dsdv{\displaystyle{\diff\sigma/\diff v}}
\def\dsdvdo{\displaystyle{\diff\sigma\over\diff v\,\diff\Omega}}
\def\dyield{\diff{\cal I}(v_{\|})}
\def\yield{\displaystyle{\dyield/\diff\vpar}}

	The physical quantity of interest in equation~(\ref{eq:equation3}) 
is the term $\dsdv$, describing the variation of the cross 
section $\sigma(v)$ as a function of the absolute velocity $v$ in the 
center-of-mass frame, whereas the measured quantity is the left-hand 
term $\yield$, representing the variation of the apparent cross section 
as a function of the longitudinal velocity component $\vpar$ in the 
center-of-mass frame.
	In principle, equation~(\ref{eq:equation3}) could not be inverted in 
an unambiguous way for general shapes of the $\dsdv$ function. 
	However, for the restricted shapes describing the data, this 
inversion becomes possible. 
	This is particularly the case if this function is supposed to 
decrease monotonically to 0 at large $v$ and if $\yield$ also follows 
the same behaviour at large $\vert\vpar\vert$, as it is evident from 
fig.~\ref{fig:fig3}.

\begin{figure}[h!]
\begin{center}
\includegraphics[angle=0, width=.8\columnwidth]{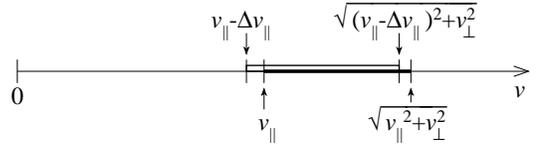}
\caption[\textheight]{
	Integration domains of eq.~(\ref{eq:equation3}).}
\label{fig:fig21}
\end{center}
\end{figure}

	The inversion procedure can be described as follows. 
	Let's consider a given bin in longitudinal velocity defined 
by the interval $[\vpar,\vpar+\Delta\vpar]$. 
	The yield for this bin is $(\yield)/\Delta\vpar$, while the 
corresponding integral over $v$ in equation~(\ref{eq:equation3}) extends 
from $\vpar$ to $\ubound{\vpar}$. 
	This domain is depicted by the thick segment in fig.~\ref{fig:fig21}.
	Let us assume that the values of the function $\dsdv$ are known over 
this interval and that they comply to equation~(\ref{eq:equation3}).
%
%

	We consider now the $\vpar$ bin located between 
$\vpar-\Delta\vpar$ and $\vpar$, for which the integration extends from  
$\vpar-\Delta\vpar$ and $\ubound{(\vpar-\Delta\vpar)}$ as shown by the thin 
segment above the axis in fig.~\ref{fig:fig21}. 
	It can be seen that this segment has a large overlap with
the previous one, provided $\Delta\vpar$ is small enough. 
	If this were not the case, the inversion procedure would even be 
simplified as $\dsdv$ would be directly proportional to the yield divided 
by the interval length, provided that the yield has a low variation over 
$\Delta\vpar$.
	In the case of overlap of the two integration segments, as shown in 
the figure, the variation of the yield comes only from the values of $\dsdv$ 
at the edges.
	If the value is known on the right non-overlapping extremity, 
the variation between the two adjacent $\vpar$ bins delivers the value on the 
left non-overlapping extremity. 
	The procedure can be continued for lower $\vpar$ bins, fixing the 
values of the function for decreasing  values of $v$.

	So far, no specific assumption has been made except that the function
$\dsdv$ is known over a given interval. 
	This can be practically achieved if one assumes that $\dsdv$
vanishes at large $v$ values and that, as a consequence, 
also the yield drops. 
	By considering a large $\vpar$ value for which the yield is null, 
we can take a null $\dsdv$ over the corresponding interval and start the  
procedure of reversion.  
	This prescription for the starting point can also be  extended to 
regions where the yield does not fade~: for $\vpar\gg\vperm$ the length of the 
integration interval decreases as $\vperm^2/(2\vpar)$, which becomes small 
compared to the characteristic variation length of $\dsdv$. In this case, 
the latter can be assumed constant over the interval and its value 
deduced straightforward from eq.~(\ref{eq:equation3}).
	The dependence of $\vperm$ on $\varphi$ only slighly changes the 
procedure, while the scheme remains the same.

	The yields measured for the forward emission ($\vpar>0$) are expected 
to differ from those associated to the backward ($\vpar<0$) emission.
	Nevertheless, in the ideal case of a perfectly isotropic emission
with respect to the center of mass, the resulting cross sections 
$\sigma (\vpar>0)$ and $\sigma (\vpar<0)$ restricted to 
only-forward and only-backward emission, respectively, should be identical.
	The difference $|\sigma (\vpar>0) - \sigma (\vpar<0)|$
can be an indication of the uncertainty introduced in the extraction of the
cross section $\sigma(v)$ by the assumption of isotropic emission.
\section{
	Isotopic cross sections		\label{appendix:cross_sections}
}
%
%
%
\begin{table}[ht!]
\caption
{
	Spallation and fragmentation residue isotopic cross sections
measured in this work for the formation of Li, Be, B, C, N and O
in the reaction $^{56}$Fe$+p$ and $^{56}$Fe+$^{\mathrm{nat}}$Ti, respectively.
}
\label{tab:tab1}
\vspace{5pt}
\begin{ruledtabular}
\begin{tabular}{c c r@{}l c r@{}l}
Isotope 
&$\;\;\;\;\;\;\;\;$&
\multicolumn{2}{l}{$^{56}$Fe$+p$, $\sigma$ [mb]} 
&$\;\;\;\;\;\;\;\;$&
\multicolumn{2}{l}{$^{56}$Fe+$^{\mathrm{nat}}$Ti, $\sigma$ [mb]} \vspace{3pt}\\
\hline\vspace{-3pt}\\
$^{6}$Li & & 14.89 &$\pm$ 1.5   & & 128.50 &$\pm$ 12.9  \\
$^{7}$Li & &  3.06 &$\pm$ 0.3   & & 103.46 &$\pm$ 10.3   \vspace{6pt}\\

$^{7}$Be & &  3.09 &$\pm$ 0.3   & &  62.28 &$\pm$ 6.2   \\
$^{9}$Be & &  2.11 &$\pm$ 0.2   & &  29.88 &$\pm$ 3.0   \vspace{6pt}\\

$^{10}$B & &  2.03 &$\pm$ 0.2   & &  35.06 &$\pm$ 3.5   \\
$^{11}$B & &  3.82 &$\pm$ 0.4   & &  61.58 &$\pm$ 6.2   \\
$^{12}$B & &  0.36 &$\pm$ 0.04  & &   9.56 &$\pm$ 1.0   \vspace{6pt}\\

$^{11}$C & &  1.12 &$\pm$ 0.1   & &  16.02 &$\pm$ 1.6   \\
$^{12}$C & &  4.69 &$\pm$ 0.4   & &  62.50 &$\pm$ 6.3   \\
$^{13}$C & &  2.76 &$\pm$ 0.3   & &  37.59 &$\pm$ 3.6   \\
$^{14}$C & &  1.87 &$\pm$ 0.2   & &  14.74 &$\pm$ 1.5   \vspace{6pt}\\
$^{13}$N & &  0.14 &$\pm$ 0.01  & &             &       \\	
$^{14}$N & &  1.25 &$\pm$ 0.1   & &  18.67 &$\pm$ 1.9   \\
$^{15}$N & &  2.97 &$\pm$ 0.3   & &  37.33 &$\pm$ 3.7   \\
$^{16}$N & &  0.33 &$\pm$ 0.03  & &   5.04 &$\pm$ 0.5   \\
$^{17}$N & &  0.11 &$\pm$ 0.01  & &   2.44 &$\pm$ 0.2   \\
$^{18}$N & &            &       & &   0.01 &$\pm$ 0.001 \vspace{6pt}\\

$^{14}$O & &  0.01 &$\pm$ 0.001 & &             &      \\
$^{15}$O & &  0.33 &$\pm$ 0.03  & &   5.91 &$\pm$ 0.6  \\
$^{16}$O & &  2.78 &$\pm$ 0.3   & &  36.62 &$\pm$ 3.7  \\
$^{17}$O & &  1.46 &$\pm$ 0.1   & &  15.58 &$\pm$ 1.5  \\
$^{18}$O & &  0.75 &$\pm$ 0.08  & &   9.12 &$\pm$ 0.9  \\
$^{19}$O & &  0.13 &$\pm$ 0.01  & &   1.99 &$\pm$ 0.2  \\
\end{tabular}
\end{ruledtabular}
\end{table}
	In the subsection~\ref{subsec:velocities} we described
how cross sections could be extracted from the measured longitudinal
velocity spectra.
	The results are presented in table~\ref{tab:tab1} with the
statistical uncertainties.
	In appendix~\ref{appendix:v_reconstr} the numerical tool
used in our analysis was presented.
	It should be observed that our velocity-reconstruction
method allows to obtain cross sections for isotopes of which
at least a half of the longitudinal velocity spectrum is measured.
	In this case, some needed parameters like the mean recoil
velocity or the width of the distribution could be extrapolated from
neighbouring isotopes.
	On the other hand, the whole procedure is valid up to a
certain extent: due to the assumption of isotropic
emission, ideal cases should result into equal (equal area and equal
centroid) absolute-velocity distributions deduced from the forward
and the backward part of the measured longitudinal velocity
distributions.
	Deviations from this ideal case derive either from the
physics of the reaction process, that could differ from a purely
isotropic emission , or by the lack of statistics in some parts of
the spectrum, resulting in complicating the convergence of the
numerical calculation.
	This leads to results that fluctuate by 10\%
in the average.
	We take this value as the statistical uncertainty (and not
simply the statistics of counts).

	The systematic uncertainties are in general very
small in FRS measurements of spallation residues.
Indeed they rise to considerably high values when the measurement
is dedicated to fragments having very high velocities in the
projectile frame.
	This is the case of very light fragments emitted in
fission-like events or in break-up processes.
	The largest source of uncertainty is the angular
acceptance. Heavy residues, close to the projectile mass are emitted
very forward, and the angular acceptance is close to
100\%.
On the contrary, light fragments are strongly affected.
	The multiplicity of the intermediate-mass fragments could not
directly be measured. From physical arguments we could safely
infer that light fragments are emitted in events with multiplicity
(of fragments with $A$>4) prevalently equal to two.
	Indeed we could not exclude the possible contribution of
higher multiplicity processes.
	We estimated the systematic uncertainty to be up to 30\%.

	It might be remarked that the greatest contribution to the
total uncertainty comes from the systematic uncertainty.
	On the other hand, the cross-section ratios of different nuclides
are very
consistent as they are related to small statistical uncertainties.
%
%
%
%
%
%

%
%
%
%
\end{document}